\documentclass[onecolumn,amsmath,amssymb,showpacs,amsfonts,a4paper,aps,prd,10pt]{revtex4}
\usepackage{graphicx}
\usepackage{epstopdf}
\newcommand{\nc}{\newcommand}
\nc{\ba}{\begin{eqnarray}} \nc{\ea}{\end{eqnarray}}
\newcommand\be{\begin{equation}}
\newcommand\ee{\end{equation}}
\nc{\D}{\overline{\mbox{D3}}}

\newcommand\vp{\varphi}
\newcommand\vpb{\bar{\varphi}}
\newcommand\chib{\bar{\chi}}
\newcommand\delvp{\delta\varphi}
\newcommand\delchi{\delta\chi}
\nc{\ga}{\gamma} \nc{\tnu}{\tilde{\nu}} \nc{\tmu}{\tilde{\mu}}

\nc{\x}{{\bf{x}}}

\input{epsf.sty}
\begin{document}

%%%%%%%%%%%%%%%%%%%%%%%%%%%%%%%%%%%%%%%%%%%%%%%%%%%%%%%%%%%%%%%%%%%%%%%
\title{Effective Field Theory of Multi-Field Inflation\\a la Weinberg}
\author{Nima Khosravi}
\email{nima@aims.ac.za}

\affiliation{Cosmology Group, African Institute for Mathematical
Sciences, Muizenberg 7945, Cape Town, South Africa}

\begin{abstract}
We employ the effective field theory approach for multi-field
inflation which is a generalization of Weinberg's work
\cite{weinberg}. In this method the first correction terms in
addition to standard terms in the Lagrangian have been considered.
These terms contain up to the fourth derivative of the fields
including the scalar field and the metric. The results show the
possible shapes of the interaction terms resulting eventually in
non-Gaussianity in a general formalism. In addition generally the
speed of sound is different but almost unity. Since in this method
the adiabatic mode is not discriminated initially so we define the
adiabatic as well as entropy modes for a specific two-field model.
It has been shown that the non-Gaussianity of the adiabatic mode and
the entropy mode are correlated in shape and amplitude. It is shown
that even for speed close to unity large non-Gaussianities are
possible in multi-field case. The amount of the non-Gaussianity
depends on the curvature of the classical path in the phase-space in
the Hubble unit such that it is large for the large curvature. In
addition it is emphasized that the time derivative of adiabatic and
entropy perturbations do not transform due to the shift symmetry as
well as the original perturbations. Though two specific combinations
of them are invariant under such a symmetry and these combinations
should be employed to construct an effective field theory of
multi-field inflation.

\end{abstract}

\maketitle \tableofcontents
%%%%%%%%%%%%%%%%%%%%%%%%%%%%%%%%%%%%%%%%%%%%%%%%%%%%%%%%%%%%%%%%%%%%%%%%%%%%%%%%%%%%%%%%%%%%
\newpage
\section{Introduction}
It is generally believed that inflation can be a solution to the
problems of standard cosmology such as the horizon, flatness and
monopole problem. In addition to these achievements, inflation's
predictions are compatible with the large scale structure and CMB
fluctuations which is strong evidence in favour of inflation. The
idea of inflation is the existence of an exponentially expanding
universe at early times. But identifying a unique theoretical
realization of this period is challenging. Many theoretical models
are compatible with the observational data. For example, they are in
agreement with adiabatic, nearly Gaussian fluctuations in the CMB
fluctuations. To potentially discriminate between them more accurate
observations, such as PLANCK, are needed. This fact is a starting
point for a huge amount of work on studying non-Gaussianity of
primordial fluctuations. In this field the effective field theory
approach to inflation has been used to study the general possible
interaction terms in the single field models
\cite{paolo,weinberg,eftsingle} and in the multi-field context
\cite{senatore,eftmulti}.

The advantage of using effective theories can be seen in two
regimes. Sometimes a full theory exists for an energy domain of
interest. In this case the effective field theory may be performed
to simplify calculations in a special sub-domain of energy. In the
second case the full theory is not known for the energy scales of
interest. Here, by imposing the symmetries of the full theory one
can still build an effective field theory. In this situation the
most general form of the allowed theory, e.g. a general Lagrangian,
is constructed; by comparison to observations unspecified
coefficients can be fixed. Eventually the deduced effective theory
may shed some lights on the real theory which is beyond our current
understanding. In the special case of inflation in addition to the
above reasons, the effective field theory approach can be used to
justify the use of scalar fields as inflatons, as well as to provide
a systematic classification of non-Gaussianities \cite{baumann}
among other properties.

In \cite{paolo} the effective field theory has been developed for an
inflationary single field model. In their approach the Lagrangian is
determined by all spatially diffeomorphism invariant operators. Then
the broken time invariance is reproduced by a scalar field which
transforms in a definite form under diffeomorphism transformation.
This scalar field is well-known as the Goldstone boson. It is shown
that this scalar field represents the curvature perturbation in the
validity regime of the effective field theory. In \cite{senatore}
the generalization to multi-field inflation is studied. The
existence of more than one field in the early universe is not
unnatural and the extra fields may have observable consequences. For
example entropy modes (a property of multi-field models) can affect
the curvature mode which is for example in the CMB. In
\cite{weinberg} an alternative approach to \cite{paolo} has been
given for the effective field theory of single field inflation. In
this approach all the possible terms containing up to the fourth
derivative of a scalar field and the metric enter the effective
Lagrangian. The final result is in agreement with \cite{paolo},
except some additional fourth ordered contributions tracing back to
geometrical terms. In this approach, due to the presence of metric
perturbations, it is possible to study the gravitational wave
behavior which differs in the propagation of waves with different
helicities.

In this work we are going to generalize Weinberg's approach
\cite{weinberg} to multi-field models. In the following we will
avoid the scalar metric perturbations by an appropriate gauge
choice. However it is mentioned that for energy scales of interest
the existence of them has no observable effects on non-Gaussianity
\cite{baumann,paolo1}. Note that in \cite{weinberg,paolo,senatore},
the additional correction terms arise via space-time derivatives of
the perturbations. However one can extend the effective field theory
formalism to include correction terms corresponding to the potential
terms. It is mentioned in \cite{baumann-green} that they have no
significant contribution to non-Gaussianities since they are highly
restricted by the effectiveness of the inflationary era. But it is
well-known that in the context of multi-field inflation the
non-Gaussianity window becomes wider and maybe observable by the
future data \cite{gpmulti}. This fact also has been considered in
the context of effective field theory for multi-field inflation in
different aspects \cite{eftmulti}.

In the next section we briefly review the main results of
\cite{weinberg}. In the third section we generalize the idea of
\cite{weinberg} to illustrate the perturbations in the most general
multi-field model. This section is based on the first appendix where
we find the most Lagrangian for multi-field models. Then in the
fourth section we concentrate on a two-field case, studying the
evolution of adiabatic and entropy modes in details. In this section
we will discuss on the amplitude and shape of non-Gaussianity in our model
and illustrate a specific example. At the end of this section we
infer to some differences between this approach and Senatore and Zaldarriaga
\cite{senatore}. In the second appendix we
compare our results with \cite{gordon} as a check. Finally we
conclude in the last section.

\section{Briefly Review of Weinberg's Approach \cite{weinberg}}
To generalize Einstein-Hilbert action in the presence of matter
field it is possible to add the terms containing higher order
derivatives in the Lagrangian in addition to the standard second
order ones. In principle these additional terms can be important in
the larger energy scales. As discussed in \cite{weinberg} this
situation occurs naturally in the inflation era before horizon
crossing. According to the observations; the Hubble parameter, $H$,
and physical momentum, $k/a$, are equal (at horizon crossing) and
much less than $M_P$ and even unification scale. But due to
denominator of physical momentum in a period before horizon crossing
the physical momentum has had larger value. As a consequence,
considering the correction terms will help us to understand better
the inflationary predictions.

Due to the above discussions Weinberg in \cite{weinberg} has studied
the effects of the fourth order derivative terms in
the Einstein-Hilbert
Lagrangian in the presence of one scalar field. We are going to
generalize this model by adding more than one scalar field which is interesting
for its well-known observational consequences. Before that let us
review very briefly\footnote{Here we report Weinberg's idea very
quickly without any details. But in the following when we are going
to study its generalization we will do it in details in Appendix \ref{appendixA}.}
the main results of \cite{weinberg}. The starting point is the
Einstein-Hilbert Lagrangian which includes the leading term
\begin{eqnarray}\label{E-H action}
{\cal{L}}_0=\sqrt{g}\bigg[-\frac{M_P^2}{2}R-\frac{M^2}{2}g^{\mu\nu}\partial_\mu
\varphi\partial_\nu\varphi-M_P^2U(\varphi)\bigg]
\end{eqnarray}
where dimensionless $\varphi\equiv\varphi_c/M$ is defined such that
the kinetic term of $\varphi_c$ has the canonical form.
Obviously $\varphi_c$ has dimension of mass. It is now easier to define
the hierarchy of different derivative terms as an advantage of introduction
of the scale $M$ explicitly\footnote{Just remember that the ${\cal{L}}$
has dimension of $M^4$ and in natural unit $\partial_\mu$ has dimension
of $M^{-1}$.}. The leading correction terms are satisfied general
covariance and contain four derivatives. These term can be reduced to
the following form
\begin{eqnarray}\label{correction terms}
\Delta{\cal{L}}=\sqrt{g}f(\vp)\bigg(g^{\mu\nu}\vp_{,\mu}\vp_{,\nu}\bigg)^2+\sqrt{g}
h_1(\vp)C^{\mu\nu\rho\sigma}C_{\mu\nu\rho\sigma}+\sqrt{g}
h_2(\vp)\varepsilon^{\kappa\lambda\mu\nu}C_{\kappa\lambda}^{\hspace{3mm}\rho\sigma}C_{\mu\nu\rho\sigma}
\end{eqnarray}
where $f$, $h_1$ and $h_2$ are some dimensionless arbitrary
functions which are assumed to be in order
one\footnote{\label{footnote7}Actually it is the second term of an
expansion with respect to the inverse of $M^2$ i.e.
``$M^2,1,M^{-2},...$". The first term is
``$-\frac{M^2}{2}g^{\mu\nu}\partial_\mu \varphi\partial_\nu\varphi$"
in (\ref{E-H action}).} and $C_{\mu\nu\rho\sigma}$ is the Weyl
tensor. It is noteworthy that the above terms are not all the
generally covariant terms containing four derivatives. But all the
allowed terms except the above terms are transformed to these terms
by employing the equation of motion for the leading term as well as
ignoring the surface terms (for details see Appendix
\ref{appendixA}). For the scalar perturbations it is convenient to
assume a gauge in which metric scalar perturbations vanish. In this
gauge by splitting the scalar field to its background and perturbed
parts as $\vp=\vpb+\delvp$ the Lagrangian becomes
\begin{eqnarray}\label{scalar-perturbation}\nonumber
{\cal{L}}&=&\sqrt{g}\bigg[-\frac{M^2}{2}g^{\mu\nu}\partial_\mu
\varphi\partial_\nu\varphi-M_P^2U(\varphi)+f(\vp)\bigg(g^{\mu\nu}\vp_{,\mu}\vp_{,\nu}\bigg)^2\bigg]\\
&=&\bar{\cal{L}}-\frac{1}{2}a^3\bigg(M^2+4f(\vpb)\dot{\vpb}^2\bigg)\times\bigg(-\dot{\delvp}^2+a^{-2}(\vec{\nabla}\delvp)^2\bigg)\\\nonumber
&+&4 a^3
f(\vpb)\dot{\vpb}^2\bigg(\dot\delvp^2+\dot\delvp^3/\dot\vpb-a^{-2}\dot\delvp(\vec{\nabla}\delvp)^2/\dot\vpb+\frac{1}{4}\dot\delvp^4/\dot\vpb^2
-\frac{1}{2}a^{-2}\dot\delvp^2(\vec{\nabla}\delvp)^2/\dot\vpb^2+\frac{1}{4}a^{-4}(\vec{\nabla}\delvp)^4/\dot\vpb^2\bigg)
\end{eqnarray}
which reduces to the Lagrangian $(19)$ in \cite{weinberg} with
$\pi\equiv\delvp/\dot\vpb$ and $\dot
H=-\dot\vpb^2(M^2+4f(\vpb)\dot{\vpb}^2)/2M_P^2$. This result is
compatible with \cite{paolo} with a minor disagreement. This
disagreement is in the presence of quartic terms as well as
quadratic and cubic terms. Due to the above Lagrangian obviously
ignoring the correction term, i.e. setting $f(\vp)=0$, results in a
model with $c_s=1$, where $c_s$ is the speed of sound. But in the
presence of the correction term the speed of sound is not one and
may cause large non-Gaussianity. In addition the terms in the third
line of (\ref{scalar-perturbation}) infer to the possible shapes of
non-Gaussianities as well as their amplitude.

In this section we very briefly reviewed the idea of \cite{weinberg}
for scalar perturbations in the context of effective field theory
for inflation. In addition to scalar perturbation in \cite{weinberg}
the tensor perturbations have been considered. It is concluded in
\cite{weinberg} that the propagation of gravitational wave depends
on the helicity of the wave in this model. In the next section we
are going to generalize the above idea for a multi-field theory of
inflation without considering the tensor perturbations. Since
existence of multi-scalar-field has no effect on the tensor
perturbations and consequently gravitational wave. The detailed
calculations for the next section is in appendix \ref{appendixA}
which is also useful for clarifying the case of one field studying
very briefly in this section.

\section{Effective Field Theory of Multi-Field Inflation}
The corresponding Lagrangian to (\ref{scalar-perturbation}) for
multi-field inflation can be written as follow, which has been deduced in
details in the appendix \ref{appendixA},
\begin{eqnarray}\label{most-general-lagrangian-simplified}
{\cal{L}}=\sqrt{g}&\bigg\{&b_3^{IJKL}(\vec\vp)\nabla_\mu\vp_I\nabla^\mu\vp_J\nabla_\nu\vp_K\nabla^\nu\vp_L-\frac{M^2}{2}\delta^{IJ}\nabla_\mu\vp_I\nabla^\mu\vp_J
-M_P^2U(\vec\vp)\\\nonumber
&+&a_1(\vec\vp)R_{\mu\nu\rho\sigma}R^{\mu\nu\rho\sigma}+a_2(\vec\vp)R_{\mu\nu}R^{\mu\nu}-\frac{M_P^2}{2}R\bigg\}
\end{eqnarray}
which exactly reduces to (\ref{scalar-perturbation}) for one field
case\footnote{\label{footnote8}In \cite{weinberg} instead of Riemann
and Ricci tensors in (\ref{most-general-lagrangian-simplified}),
Weyl tensor has been used.}. Now, by splitting the scalar fields to
their background and perturbed parts $\vp_I=\bar\vp_I+\delvp$ we are
going to study the Lagrangian for the perturbations as well as the
background. To do this we start with
(\ref{most-general-lagrangian-simplified}) without worrying about
the tensor perturbations. The above Lagrangian can be written as
${\cal{L}}={\cal{L}}_0+\Delta{\cal{L}}$ such that
\begin{eqnarray}\label{most-general-lagrangian-simplified+background}\nonumber
a^{-3}{\cal{L}}_0&=&b_3^{IJKL}(\vpb){\dot\vpb}_I{\dot\vpb}_J{\dot\vpb}_K{\dot{\bar\vp}}_L
+\frac{M^2}{2}\delta^{IJ}{\dot\vpb}_I{\dot\vpb}_J-M_P^2U(\vpb),
\end{eqnarray}
\begin{eqnarray}\label{most-general-lagrangian-simplified+perturbations}\nonumber
a^{-3}\Delta{\cal{L}}&=&\bigg[\sum_{n=1}\frac{1}{n!}\frac{\partial^n
b_3^{IJKL}(\vpb)}{\partial(\vpb_M)^n}(\delvp_M)^n\bigg]{\dot\vpb}_I{\dot\vpb}_J{\dot\vpb}_K{\dot{\bar\vp}}_L\\\nonumber
&-& \bigg[\sum_{n=0}\frac{1}{n!}\frac{\partial^n
b_3^{IJKL}(\vpb)}{\partial(\vpb_M)^n}(\delvp_M)^n\bigg]\dot\vpb_I\dot\vpb_J\bigg[-\dot\vpb_K\dot\delvp_L-\dot\vpb_L\dot\delvp_K-\dot\delvp_K\dot\delvp_L+
a^{-2}\partial_i\delvp_K\partial^i\delvp_L\bigg]
\\\nonumber
&-& \bigg[\sum_{n=0}\frac{1}{n!}\frac{\partial^n
b_3^{IJKL}(\vpb)}{\partial(\vpb_M)^n}(\delvp_M)^n\bigg]\bigg[-\dot\vpb_I\dot\delvp_J-\dot\vpb_J\dot\delvp_I-\dot\delvp_I\dot\delvp_J+
a^{-2}\partial_i\delvp_I\partial^i\delvp_J\bigg]\dot\vpb_K\dot\vpb_L\\\nonumber
&+&\bigg[\sum_{n=0}\frac{1}{n!}\frac{\partial^n
b_3^{IJKL}(\vpb)}{\partial(\vpb_M)^n}(\delvp_M)^n\bigg]\times\\\nonumber&&\bigg[-\dot\vpb_I\dot\delvp_J-\dot\vpb_J\dot\delvp_I-\dot\delvp_I\dot\delvp_J+
a^{-2}\partial_i\delvp_I\partial^i\delvp_J\bigg]\bigg[-\dot\vpb_K\dot\delvp_L-\dot\vpb_L\dot\delvp_K-\dot\delvp_K\dot\delvp_L+
a^{-2}\partial_i\delvp_K\partial^i\delvp_L\bigg]\\\nonumber
&-&\frac{M^2}{2}\delta^{IJ}\bigg[-2\dot\vpb_I\dot\delvp_J-\dot\delvp_I\dot\delvp_J+
a^{-2}\partial_i\delvp_I\partial^i\delvp_J\bigg]
-M_P^2\bigg[\sum_{n=1}\frac{1}{n!}\frac{\partial^n
U(\vpb)}{\partial(\vpb_M)^n}(\delvp_M)^n\bigg]
\end{eqnarray}
where $a=a(t)$ is the scale factor of the FRW metric and
$\partial_i$ are spatial derivatives. Note that the terms containing
$\delvp_I$ without any differentiations do not show themselves in
the Lagrangian effectively. Since the $n^{th}$ equation of motion
causes vanishing of the coefficients of $(n+1)^{th}$ terms without
any differentiation. Also the linear perturbation terms even
including differentiation vanish because of the same reason. So
effectively the Lagrangian for the perturbations is
\begin{eqnarray}\label{most-general-lagrangian-simplified+perturbations-effectively}\nonumber
a^{-3}\Delta{\cal{L}}&=&
b_3^{IJKL}(\vpb)\bigg\{\bigg[-\dot\vpb_I\dot\delvp_J-\dot\vpb_J\dot\delvp_I-\dot\delvp_I\dot\delvp_J+
a^{-2}\partial_i\delvp_I\partial^i\delvp_J\bigg]\bigg[-\dot\vpb_K\dot\delvp_L-\dot\vpb_L\dot\delvp_K-\dot\delvp_K\dot\delvp_L+
a^{-2}\partial_i\delvp_K\partial^i\delvp_L\bigg]\\\nonumber
&-&\dot\vpb_I\dot\vpb_J\bigg[-\dot\delvp_K\dot\delvp_L+
a^{-2}\partial_i\delvp_K\partial^i\delvp_L\bigg]-\bigg[-\dot\delvp_I\dot\delvp_J+
a^{-2}\partial_i\delvp_I\partial^i\delvp_J\bigg]\dot\vpb_K\dot\vpb_L\bigg\}\\\nonumber
&-&\frac{M^2}{2}\delta^{IJ}\bigg[-\dot\delvp_I\dot\delvp_J+
a^{-2}\partial_i\delvp_I\partial^i\delvp_J\bigg].
\end{eqnarray}
The second, third and fourth order of perturbations respectively can
be represented as follows
\begin{eqnarray}\label{most-general-lagrangian-simplified+perturbations-effectively-L2}\nonumber
a^{-3}\Delta{\cal{L}}^{(2)}&=&
b_3^{IJKL}(\vpb)\bigg\{\dot\vpb_I\dot\vpb_J\dot\delvp_K\dot\delvp_L+\dot\vpb_I\dot\vpb_K\dot\delvp_J\dot\delvp_L+
\dot\vpb_I\dot\vpb_L\dot\delvp_K\dot\delvp_J+\dot\vpb_K\dot\vpb_J\dot\delvp_I\dot\delvp_L+
\dot\vpb_L\dot\vpb_J\dot\delvp_I\dot\delvp_L+\dot\vpb_K\dot\vpb_L\dot\delvp_I\dot\delvp_J\\&-&
a^{-2}\dot\vpb_I\dot\vpb_J\partial_i\delvp_K\partial^i\delvp_L-a^{-2}\dot\vpb_K\dot\vpb_L\partial_i\delvp_I\partial^i\delvp_J
\bigg\}-\frac{M^2}{2}\delta^{IJ}\bigg[-\dot\delvp_I\dot\delvp_J+
a^{-2}\partial_i\delvp_I\partial^i\delvp_J\bigg],
\end{eqnarray}
\begin{eqnarray}\label{most-general-lagrangian-simplified+perturbations-effectively-L3}
a^{-3}\Delta{\cal{L}}^{(3)}&=&
b_3^{IJKL}(\vpb)\bigg\{\dot\vpb_I\dot\delvp_J\dot\delvp_K\dot\delvp_L+\dot\vpb_J\dot\delvp_I\dot\delvp_K\dot\delvp_L
+\dot\vpb_K\dot\delvp_L\dot\delvp_I\dot\delvp_J+\dot\vpb_L\dot\delvp_K\dot\delvp_I\dot\delvp_J\\\nonumber
&-&a^{-2}\bigg(\dot\vpb_I\dot\delvp_J\partial_i\delvp_K\partial^i\delvp_L+\dot\vpb_J\dot\delvp_I\partial_i\delvp_K\partial^i\delvp_L
+\dot\vpb_K\dot\delvp_L\partial_i\delvp_I\partial^i\delvp_J+\dot\vpb_L\dot\delvp_K\partial_i\delvp_I\partial^i\delvp_J\bigg)\bigg\},\\\nonumber&&
\end{eqnarray}
\begin{eqnarray}\label{most-general-lagrangian-simplified+perturbations-effectively-L4}
a^{-3}\Delta{\cal{L}}^{(4)}=
b_3^{IJKL}(\vpb)\bigg\{\dot\delvp_I\dot\delvp_J\dot\delvp_K\dot\delvp_L&-&a^{-2}\bigg(\dot\delvp_I\dot\delvp_J\partial_i\delvp_K\partial^i\delvp_L+
\dot\delvp_K\dot\delvp_L\partial_i\delvp_I\partial^i\delvp_J\bigg)\\\nonumber
&+&a^{-4}\partial_i\delvp_I\partial^i\delvp_J\partial_j\delvp_K\partial^j\delvp_L\bigg\}.
\end{eqnarray}
Note that the above result exactly reduces to single field result in
(\ref{scalar-perturbation}) with
$b_3^{IJKL}(\vpb)=f(\vpb)$\footnote{To do this one should set
$I=J=K=L=1$.}. It is obvious from
(\ref{most-general-lagrangian-simplified+perturbations-effectively-L2})
that the speed of sound is not one in the presence of non-vanishing
$b_3^{IJKL}(\vpb)$. Note that due to $b_3^{IJKL}(\vpb)$ the cubic
and quartic terms are appeared. This feature is in disagreement with
\cite{paolo,senatore}. In their work the coefficient which displays
$c_s$ just connects to the cubic term. But here it connects to the
fourth order term too. In the next section we restrict the model to
a two-field model. This makes it possible to study the adiabatic and
entropy perturbations in more details without loss of generality in
the main results.

\section{A Specific Case: Adiabatic versus Entropy Perturbation}
In this section we re-write the above formalism in the language of
adiabatic and entropy perturbations for a two-field model. This is
crucial in this approach since in contrast to \cite{senatore} here
the adiabatic perturbation is not initially supposed. In
\cite{senatore} the additional perturbations are added to a model
already containing the adiabatic perturbation i.e. \cite{paolo}. In
\cite{senatore} the Goldstone boson, introduced in \cite{paolo},
plays the role of the adiabatic perturbation and the additional
fields are employed as the entropy perturbations. But in our model
there is no initially difference between $\vp_I$'s and consequently
$\delvp_I$'s. So it is critical to distinguish between adiabatic and
entropy modes to manifest their different physical meanings.

\subsection{The Most General Two-Field Model}
In this subsection we re-do perturbation calculations for a two-field model.
To do this we start
with (\ref{most-general-lagrangian-simplified}) for two fields named
$\vp$ and $\chi$
\begin{eqnarray}\label{two-field-lagrangian-simplified}
{\cal{L}}=-a^3\bigg\{&-&\frac{M_1^2}{2}\partial_\mu\vp\partial^\mu\vp-\frac{M_2^2}{2}\partial_\mu\chi\partial^\mu\chi
-M_P^2U(\vp,\chi)+g_1(\vp,\chi)\big(\partial_\mu\vp\partial^\mu\vp\big)^2+g_2(\vp,\chi)\big(\partial_\mu\chi\partial^\mu\chi\big)^2\\\nonumber
&+&g_3(\vp,\chi)\big(\partial_\mu\vp\partial^\mu\vp\big)\big(\partial_\nu\vp\partial^\nu\chi\big)
+g_4(\vp,\chi)\big(\partial_\mu\chi\partial^\mu\chi\big)\big(\partial_\nu\chi\partial^\nu\vp\big)\\\nonumber&+&g_5(\vp,\chi)\big(\partial_\mu\vp\partial^\mu\vp\big)\big(\partial_\nu\chi\partial^\nu\chi\big)
+g_6(\vp,\chi)\big(\partial_\mu\vp\partial^\mu\chi\big)\big(\partial_\nu\vp\partial^\nu\chi\big)\bigg\}
\end{eqnarray}
where $a=a(t)$ is the scale factor and $g_i$'s are some arbitrary
dimensionless and order one functions as mentioned before. By assuming
$\vp=\vpb+\delvp$ and $\chi=\bar\chi+\delchi$ the above Lagrangian
reduces to
\begin{eqnarray}\label{two-field-lagrangian-simplified-order-0}
a^{-3}{\cal{L}}_0&=&\bigg\{-\frac{M_1^2}{2}\partial_\mu\vpb\partial^\mu\vpb-\frac{M_2^2}{2}\partial_\mu\chib\partial^\mu\chib
-M_P^2U(\vpb,\chib)+g_1\big(\partial_\mu\vpb\partial^\mu\vpb\big)^2+g_2\big(\partial_\mu\chib\partial^\mu\chib\big)^2
\\\nonumber&+&g_3\big(\partial_\mu\vpb\partial^\mu\vpb\big)\big(\partial_\nu\vpb\partial^\nu\chib\big)
+g_4\big(\partial_\mu\chib\partial^\mu\chib\big)\big(\partial_\nu\chib\partial^\nu\vpb\big)+
g_5\big(\partial_\mu\vpb\partial^\mu\vpb\big)\big(\partial_\nu\chib\partial^\nu\chib\big)
+g_6\big(\partial_\mu\vpb\partial^\mu\chib\big)\big(\partial_\nu\vpb\partial^\nu\chib\big)\bigg\}\\\nonumber
&=&\frac{M_1}{2}\dot\vpb^2+\frac{M_2}{2}\dot\chib^2-M_P^2U(\vpb,\chib)+g_1\dot\vpb^4+g_2\dot\chib^4+g_3\dot\vpb^3\dot\chib+g_4\dot\vpb\dot\chib^3
+(g_5+g_6)\dot\vpb^2\dot\chib^2
\end{eqnarray}
for the background part. The corresponding equations of motion for $\vpb$ reads as
\begin{eqnarray}\label{eq-mo-background}\nonumber
&&\frac{d}{dt}\bigg[M_1\dot\vpb+4g_1\dot\vpb^3+3g_3\dot\vpb^2\dot\chib+g_4\dot\chib^3
+2(g_5+g_6)\dot\vpb\dot\chib^2\bigg]+3H\bigg[M_1\dot\vpb+4g_1\dot\vpb^3+3g_3\dot\vpb^2\dot\chib+g_4\dot\chib^3
+2(g_5+g_6)\dot\vpb\dot\chib^2\bigg]\\\nonumber
&&+M_P^2U'-\bigg(g'_1\dot\vpb^4+g'_2\dot\chib^4+g'_3\dot\vpb^3\dot\chib+g'_4\dot\vpb\dot\chib^3
+(g'_5+g'_6)\dot\vpb^2\dot\chib^2\bigg)=0
\end{eqnarray}
where $'$ is the differentiation with respect to $\vpb$ and the
similar equation is true for $\chib$. It is straightforward but
messy to show that the above equation of motion (as well as
$\chib$'s) for the background causes the Lagrangian of the first
order perturbation becomes vanishing. So the non-trivial terms start
from the second order perturbations succeeding with the third and
the fourth order terms\footnote{It is obvious if one expands the
correction terms in (\ref{most-general-lagrangian-simplified}) or
(\ref{most-general-lagrangian}) for more than four derivative terms
then the higher order perturbations show themselves.} for
(\ref{two-field-lagrangian-simplified}) as the following
\begin{eqnarray}\label{two-field-lagrangian-simplified-order-2}
a^{-3}\Delta{\cal{L}}^{(2)}&=&\dot\delvp^2\big[\frac{M^2_1}{2}+6g_1\dot\vpb^2+3g_3\dot\vpb\dot\chib+(g_5+g_6)\dot\chib^2\big]
+\dot\delchi^2\big[\frac{M^2_2}{2}+6g_2\dot\chib^2+3g_4\dot\vpb\dot\chib+(g_5+g_6)\dot\vpb^2\big]\\\nonumber&+&
\dot\delvp\dot\delchi\big[3g_3\dot\vpb^2+3g_4\dot\chib^2+4(g_5+g_6)\dot\vpb\dot\chib\big]\\\nonumber
&-&a^{-2}\bigg(\partial_i\delvp\partial^i\delvp\big[\frac{M_1}{2}+2g_1\dot\vpb^2+g_3\dot\vpb\dot\chib+g_5\dot\chib^2\big]
+\partial_i\delchi\partial^i\delchi\big[\frac{M_2}{2}+2g_2\dot\chib^2+g_4\dot\vpb\dot\chib+g_5\dot\vpb^2\big]\\\nonumber
&&\hspace{1.2cm}+\partial_i\delvp\partial^i\delchi\big[g_3\dot\vpb^2+g_4\dot\chib^2+2g_6\dot\vpb\dot\chib\big]\bigg),
\end{eqnarray}
\begin{eqnarray}\label{two-field-lagrangian-simplified-order-3}
a^{-3}\Delta{\cal{L}}^{(3)}&=&\dot\delvp^3\big[4g_1\dot\vpb+g_3\dot\chib\big]+\dot\delchi^3\big[4g_2\dot\chib+g_4\dot\vpb\big]+
\dot\delvp^2\dot\delchi\big[3g_3\dot\vpb+2(g_5+g_6)\dot\chib\big]+\dot\delvp\dot\delchi^2\big[3g_4\dot\chib+2(g_5+g_6)\dot\vpb\big]\\\nonumber
&-&a^{-2}\bigg(\dot\delvp\partial_i\delvp\partial^i\delvp\big[4g_1\dot\vpb+g_3\dot\chib\big]+
\dot\delchi\partial_i\delchi\partial^i\delchi\big[4g_2\dot\chib+g_4\dot\vpb\big]+
\dot\delvp\partial_i\delchi\partial^i\delchi\big[g_4\dot\chib+2g_5\dot\vpb\big]\\\nonumber&&\hspace{1.5cm}
+\dot\delchi\partial_i\delvp\partial^i\delvp\big[g_3\dot\vpb+2g_5\dot\chib\big]
+\dot\delvp\partial_i\delvp\partial^i\delchi\big[2g_3\dot\vpb+2g_6\dot\chib\big]+
\dot\delchi\partial_i\delvp\partial^i\delchi\big[2g_4\dot\chib+2g_6\dot\vpb\big]\bigg),
\end{eqnarray}
\begin{eqnarray}\label{two-field-lagrangian-simplified-order-4}
a^{-3}\Delta{\cal{L}}^{(4)}&=&g_1\dot\delvp^4+g_2\dot\delchi^4+g_3\dot\delvp^3\dot\delchi+g_4\dot\delchi^3\dot\delvp+(g_5+g_6)\dot\delvp^2\dot\delchi^2\\\nonumber
&-&a^{-2}\bigg(2g_1\dot\delvp^2\partial_i\delvp\partial^i\delvp+2g_2\dot\delchi^2\partial_i\delchi\partial^i\delchi+
g_3\dot\delvp^2\partial_i\delvp\partial^i\delchi+g_4\dot\delchi^2\partial_i\delchi\partial^i\delvp\\\nonumber
&&\vspace{2.5cm}+g_3\dot\delvp\dot\delchi\partial_i\delvp\partial^i\delvp+g_4\dot\delchi\dot\delvp\partial_i\delchi\partial^i\delchi+
g_5(\dot\delchi^2\partial_i\delvp\partial^i\delvp+\dot\delvp^2\partial_i\delchi\partial^i\delchi)+2g_6\dot\delvp
\dot\delchi\partial_i\delvp\partial^i\delchi\bigg)\\\nonumber
&+&a^{-4}\bigg(g_1(\partial_i\delvp\partial^i\delvp)^2+g_2(\partial_i\delchi\partial^i\delchi)^2
+g_3\partial_i\delvp\partial^i\delvp\partial_j\delchi\partial^j\delvp+g_4\partial_i\delchi\partial^i\delchi\partial_j\delchi\partial^j\delvp
\\\nonumber&&\vspace{2.5cm}+g_5\partial_i\delvp\partial^i\delvp\partial_j\delchi\partial^j\delchi+g_6(\partial_i\delvp\partial^i\delchi)^2\bigg).
\end{eqnarray}
It is interesting to mention that for $\vp=\chi$, i.e. going back to one
field case, all the above relations reduce to
(\ref{scalar-perturbation}) with $f=g_1+g_2+g_3+g_4+g_5+g_6$ as it was expected.

\subsection{Adiabatic vs. Entropy Modes}
Now let us re-write the above terms in the language of adiabatic and
entropy perturbations. The adiabatic perturbation is along the classical
path and the entropy perturbation
is orthogonal to it. Due to \cite{gordon} they can be defined as
follows
\begin{eqnarray}\label{adi-ent-perturbations}
\delta\sigma\equiv\vec{T}.\vec\delta,\hspace{2cm}\delta s\equiv
\vec{N}.\vec\delta
\end{eqnarray}
where $\delta \sigma$ and $\delta s$ are the adiabatic and entropy
modes respectively and
\begin{eqnarray}\label{tangent-normal-vector}
\vec\delta\equiv\left(\delvp,\delchi\right),\hspace{2cm}\vec
T=\left(\cos\theta,\sin\theta\right)\equiv\left(\dot\vp/\dot\sigma,\dot\chi/\dot\sigma\right),\hspace{2cm}
\vec N\equiv\left(\sin\theta,-\cos\theta\right)
\end{eqnarray}
where $\dot\sigma^2=\dot\vp^2+\dot\chi^2$. One can show easily
\begin{eqnarray}\label{adi-ent-perturbations-timederivative-original-perturbations}\nonumber
\dot\delvp=\cos\theta\hspace{2mm}\vec{T}.\dot{\vec\delta}+\sin\theta\hspace{2mm}\vec{N}.\dot{\vec\delta},\hspace{2cm}
\dot\delchi=\sin\theta\hspace{2mm}\vec{T}.\dot{\vec\delta}-\cos\theta\hspace{2mm}\vec{N}.\dot{\vec\delta}
\end{eqnarray}
and it is easy to see that
\begin{eqnarray}\label{adi-ent-perturbations-timederivative}\nonumber
\dot{\delta\sigma}=\dot\theta\delta
s+\vec{T}.\dot{\vec\delta},\hspace{2cm}\dot{\delta s}=-\dot\theta
\delta\sigma+ \vec{N}.\dot{\vec\delta}
\end{eqnarray}
and due to above relations
\begin{eqnarray}\label{adi-ent-perturbations-spatialderivative-original-perturbations}\nonumber
\partial_i\delvp=\cos\theta\hspace{2mm}\partial_i\delta\sigma+\sin\theta\hspace{2mm}\partial_i\delta s,\hspace{2cm}
\partial_i\delchi=\sin\theta\hspace{2mm}\partial_i\delta\sigma-\cos\theta\hspace{2mm}\partial_i\delta
s.
\end{eqnarray}
Now by the above definitions we re-write the results of previous
sub-section by plugging $\delta\sigma$ and $\delta s$ in. Let's
start with the kinetic terms for leading order term in
(\ref{two-field-lagrangian-simplified-order-2})
\begin{eqnarray}\label{two-field-lagrangian-simplified-order-2-M1-M2-adi-ent}
&&\frac{M^2_1}{2}\dot\delvp^2+\frac{M^2_2}{2}\dot\delchi^2=\\\nonumber
&&\frac{M^2_1+M_2^2}{4}\left[\dot\delvp^2+\dot\delchi^2\right]+\frac{M_1^2-M^2_2}{4}\left[\dot\delvp^2-\dot\delchi^2\right]=\\\nonumber&&
\frac{M^2_1+M_2^2}{4}\left[(\vec T.\dot{\vec\delta})^2+(\vec
N.\dot{\vec\delta})^2\right]+\frac{M_1^2-M^2_2}{4}\left[(\cos^2\theta-\sin^2\theta)\bigg((\vec
T.\dot{\vec\delta})^2-(\vec
N.\dot{\vec\delta})^2\bigg)+4\sin\theta\cos\theta\hspace{2mm}
T.\dot{\vec\delta}\hspace{2mm}N.\dot{\vec\delta}\right]\\\nonumber&&
\end{eqnarray}
the same procedure is applicable for
$a^{-2}\left(\frac{M^2_1}{2}\partial_i\delvp\partial^i\delvp
+\frac{M^2_2}{2}\partial_i\delchi\partial^i\delchi\right)$ in
(\ref{two-field-lagrangian-simplified-order-2}). Now let us assume
$M_1=M_2=M$ to make it comparable to results in \cite{gordon}. In
this case
\begin{eqnarray}\label{two-field-lagrangian-simplified-order-2-M1=M2=1-adi-ent}
\frac{1}{M^2}a^{-3}{\cal{L}}&=&\frac{1}{2}\dot\delvp^2+\frac{1}{2}\dot\delchi^2-a^{-2}\left(\frac{1}{2}\partial_i\delvp\partial^i\delvp
+\frac{1}{2}\partial_i\delchi\partial^i\delchi\right)=\\\nonumber &&
\frac{1}{2}\left[(\vec T.\dot{\vec\delta})^2+(\vec
N.\dot{\vec\delta})^2\right]-\frac{1}{2}a^{-2}\left(\partial_i\delta\sigma\partial^i\delta\sigma
+\partial_i\delta s\partial^i\delta s\right)=\\\nonumber&&
 \frac{1}{2}\left[(\dot{\delta\sigma}-\dot\theta\delta s)^2+(\dot{\delta s}+\dot\theta\delta\sigma)^2\right]
 -\frac{1}{2}a^{-2}\left(\partial_i\delta\sigma\partial^i\delta\sigma
+\partial_i\delta s\partial^i\delta s\right)=\\\nonumber&&
 \frac{1}{2}\left[\dot{\delta\sigma}^2+\dot\theta^2\delta s^2-2\dot\theta\delta s\dot{\delta\sigma}+\dot{\delta s}^2
 +\dot\theta^2\delta\sigma^2+2\dot\theta\delta\sigma\dot{\delta s}\right]-\frac{1}{2}a^{-2}\left(\partial_i\delta\sigma\partial^i\delta\sigma
+\partial_i\delta s\partial^i\delta s\right)
\end{eqnarray}
According to the above Lagrangian the equations of motion
for $\delta\sigma$ and $\delta s$ become\footnote{The potential term in the second order of
perturbations should be added to
(\ref{two-field-lagrangian-simplified-order-2-M1=M2=1-adi-ent}) to
make our results comparable with \cite{gordon}. This term is
$-\frac{1}{2}\big(V_{\sigma\sigma}\delta\sigma^2+V_{\sigma s}\delta
\sigma \delta s+V_{ss}\delta s^2\big)$.}
\begin{eqnarray}\label{eq-mo-pert}
&&\ddot{\delta\sigma}+3H\dot{\delta\sigma}-a^{-2}\partial^i\partial_i\delta\sigma+(V_{\sigma\sigma}-\dot\theta^2)\delta\sigma=2\dot\theta\dot{\delta
s} +(\ddot\theta+3H\dot\theta-V_{\sigma s})\delta s\\\nonumber
&&\ddot{\delta s}+3H\dot{\delta s}-a^{-2}\partial^i\partial_i\delta
s+(V_{ss}-\dot\theta^2)\delta s=-2\dot\theta\dot{\delta \sigma}
-(\ddot\theta+3H\dot\theta+V_{\sigma s})\delta \sigma
\end{eqnarray}
where $V_{\sigma s}=(\cos^2\theta-\sin^2\theta)
V_{\vp\chi}+\sin\theta\cos\theta(V_{\chi\chi}-V_{\vp\vp})$. The
above results are exactly same as (47) and (48) in \cite{gordon}
when ignoring metric perturbations, see Appendix B. Now let us do
the same procedure for the second order perturbations due to the
first order correction term in
(\ref{two-field-lagrangian-simplified-order-2}). At the first the
terms containing time derivative
\begin{eqnarray}\label{adi-ent-pert-correction-order-2}\nonumber
&&\dot\delvp^2\big[6g_1\dot\vpb^2+3g_3\dot\vpb\dot\chib+(g_5+g_6)\dot\chib^2\big]+
\dot\delchi^2\big[6g_2\dot\chib^2+3g_4\dot\vpb\dot\chib+(g_5+g_6)\dot\vpb^2\big]+
\dot\delvp\dot\delchi\big[3g_3\dot\vpb^2+3g_4\dot\chib^2+4(g_5+g_6)\dot\vpb\dot\chib\big]\\\nonumber&=&
\bigg(\cos\theta\hspace{2mm}\vec{T}.\dot{\vec\delta}+\sin\theta\hspace{2mm}\vec{N}.\dot{\vec\delta}\bigg)^2
\bigg[6g_1\dot\vpb^2+3g_3\dot\vpb\dot\chib+(g_5+g_6)\dot\chib^2\bigg]+
\bigg(\sin\theta\hspace{2mm}\vec{T}.\dot{\vec\delta}-\cos\theta\hspace{2mm}\vec{N}.\dot{\vec\delta}\bigg)^2
\bigg[6g_2\dot\chib^2+3g_4\dot\vpb\dot\chib+(g_5+g_6)\dot\vpb^2\bigg]\\\nonumber&+&
\bigg(\cos\theta\hspace{2mm}\vec{T}.\dot{\vec\delta}+\sin\theta\hspace{2mm}\vec{N}.\dot{\vec\delta}\bigg)
\bigg(\sin\theta\hspace{2mm}\vec{T}.\dot{\vec\delta}-\cos\theta\hspace{2mm}\vec{N}.\dot{\vec\delta}\bigg)
\bigg[3g_3\dot\vpb^2+3g_4\dot\chib^2+4(g_5+g_6)\dot\vpb\dot\chib\bigg]\\\nonumber&=&
\big(\vec{T}.\dot{\vec\delta}\big)^2\times\bigg\{\cos^2\theta\bigg[6g_1\dot\vpb^2+3g_3\dot\vpb\dot\chib+(g_5+g_6)\dot\chib^2\bigg]
+\sin^2\theta\bigg[6g_2\dot\chib^2+3g_4\dot\vpb\dot\chib+(g_5+g_6)\dot\vpb^2\bigg]\\\nonumber&&\hspace{2cm}+\sin\theta\cos\theta
\bigg[3g_3\dot\vpb^2+3g_4\dot\chib^2+4(g_5+g_6)\dot\vpb\dot\chib\bigg]\bigg\}\\\nonumber&+&
\big(\vec{N}.\dot{\vec\delta}\big)^2\times\bigg\{\sin^2\theta\bigg[6g_1\dot\vpb^2+3g_3\dot\vpb\dot\chib+(g_5+g_6)\dot\chib^2\bigg]
+\cos^2\theta\bigg[6g_2\dot\chib^2+3g_4\dot\vpb\dot\chib+(g_5+g_6)\dot\vpb^2\bigg]\\\nonumber&&\hspace{2cm}-\sin\theta\cos\theta
\bigg[3g_3\dot\vpb^2+3g_4\dot\chib^2+4(g_5+g_6)\dot\vpb\dot\chib\bigg]\bigg\}\\\nonumber&+&
\big(\vec{T}.\dot{\vec\delta}\big)\big(\vec{N}.\dot{\vec\delta}\big)\times
\bigg\{2\cos\theta\sin\theta\bigg(\big[6g_1\dot\vpb^2+3g_3\dot\vpb\dot\chib+(g_5+g_6)\dot\chib^2\big]
-\big[6g_2\dot\chib^2+3g_4\dot\vpb\dot\chib+(g_5+g_6)\dot\vpb^2\big]\bigg)\\\nonumber&&\hspace{2.5cm}+\bigg(\sin^2\theta-\cos^2\theta\bigg)
\bigg[3g_3\dot\vpb^2+3g_4\dot\chib^2+4(g_5+g_6)\dot\vpb\dot\chib\bigg]\bigg\}
\end{eqnarray}
which can be re-written as the following\footnote{Here we do not
expand $\vec{T}.\dot{\vec\delta}$ and $\vec{N}.\dot{\vec\delta}$
since they contain no common terms to factorize. So their expansion
may cause just messy stuffs without any physical interests.}
\begin{eqnarray}\label{two-field-lagrangian-simplified-order-2-correction-terms}
&&6\dot\sigma^2\big(\vec{T}.\dot{\vec\delta}\big)^2\times\bigg[g_1\cos^4\theta+g_2\sin^4\theta+g_3\cos^3\theta\sin\theta
+g_4\cos\theta\sin^3\theta+(g_5+g_6)\cos^2\theta\sin^2\theta\bigg]\\\nonumber&+&
\dot\sigma^2\big(\vec{N}.\dot{\vec\delta}\big)^2\times\\\nonumber&&\vspace{0cm}\bigg[(g_5+g_6)\bigg(\cos^4\theta+\sin^4\theta\bigg)+3(g_4-g_3)\bigg(\cos^3\theta\sin\theta
-\cos\theta\sin^3\theta\bigg)+2\bigg(3(g_1+g_2)+2(g_5+g_6)\bigg)\cos^2\theta\sin^2\theta\bigg]\\\nonumber&+&
3\dot\sigma^2\big(\vec{T}.\dot{\vec\delta}\big)\big(\vec{N}.\dot{\vec\delta}\big)\times\\\nonumber&&
\bigg[-g_3\cos^4\theta+g_4\sin^4\theta+2\bigg(2g_1-(g_5+g_6)\bigg)\cos^3\theta\sin\theta-2\bigg(2g_2+(g_5+g_6)\bigg)\cos\theta\sin^3\theta
+3(g_3-g_4)\cos^2\theta\sin^2\theta\bigg]
\end{eqnarray}
and similarly for the spatial differentiation
\begin{eqnarray}\label{two-field-lagrangian-simplified-order-2-correction-terms-spatial}
&-&2a^{-2}\dot\sigma^2(\partial_i\delta\sigma)^2\times\bigg[g_1\cos^4\theta+g_2\sin^4\theta+g_3\cos^3\theta\sin\theta
+g_4\cos\theta\sin^3\theta+2(g_5+g_6)\cos^2\theta\sin^2\theta\bigg]\\\nonumber&-&
a^{-2}\dot\sigma^2(\partial_i\delta
s)^2\times\vspace{0cm}\bigg[g_5\bigg(\cos^4\theta+\sin^4\theta\bigg)+(g_4-g_3)\bigg(\cos^3\theta\sin\theta
-\cos\theta\sin^3\theta\bigg)+2(g_1+g_2-g_6)\cos^2\theta\sin^2\theta\bigg]\\\nonumber&-&
a^{-2}\dot\sigma^2\partial_i\delta\sigma\partial^i \delta
s\times\\\nonumber&&
\bigg[-g_3\cos^4\theta+g_4\sin^4\theta+2\bigg(2g_1-(g_5+g_6)\bigg)\cos^3\theta\sin\theta-2\bigg(2g_2+(g_5+g_6)\bigg)\cos\theta\sin^3\theta
+3(g_3-g_4)\cos^2\theta\sin^2\theta\bigg].
\end{eqnarray}
Up to now we fully considered the second order perturbation terms in
the language of adiabatic and entropy modes. In quadratic level the
speed of sound is a matter of interest hence it is noteworthy to
take a look at it. It is obvious to the above relations that $\delta
\sigma$ and $\delta s$ can have different speeds of sound generally.
To have a sense about it let's consider a special case that $\delta
\sigma$ has $c_s=1$ and $\delta s$ has $c_s\neq 1$. To see this,
assume the special case with $g_1=g_2=g_3=g_4=0$ and $g_5+g_6=0$. In
this case the coefficients of $\dot{\delta\sigma}^2$ and
$(\partial_i\delta\sigma)^2$ are same and result in $c_s=1$ for
$\delta\sigma$\footnote{Note that we employ the standard definition
of $c_s$. It means we skip the interaction terms between
$\delta\sigma$ and $\delta s$ which exist even in quadratic level.}.
But the coefficient of $\dot{\delta s}^2$ is $\frac{M^2}{2}$ and for
$(\partial_i\delta s)^2$ is
$-\frac{1}{2}a^{-2}(M^2-2g_5\dot\sigma^2)$ that means
$c_s^2=1-2\frac{g_5\dot\sigma^2}{M^2}$. Note that here we write the
$M$ explicitly to make the comparison of the terms easier. A
characteristic property of $c_s$ is its $\dot\sigma^2$-dependence
which seems interesting. However the effective field theory is valid
where the correction terms are smaller than the leading terms in
(\ref{most-general-lagrangian-simplified}) to have an acceptable
expansion i.e. $\frac{\vert
b_3^{IJKL}(\vpb)\vert\dot{\sigma}^2}{M^2}<1$. Even more,
$\frac{\vert b_3^{IJKL}(\vpb)\vert\dot{\sigma}^2}{M^2}<<1$ should be
satisfied to make skipping higher order correction terms in
(\ref{most-general-lagrangian-simplified}) acceptable. So the speed
of sound in this model is almost one. This fact shows that for the
single field model the large non-Gaussinity is not expected. But in
the following we will discuss on the case of multi-field models. In
multi-field models even with $c_s\simeq 1$ the large non-Gaussianity
can be occurred in some specific circumstances.

One can do this procedure for the higher order perturbation terms
(\ref{two-field-lagrangian-simplified-order-3}) and
(\ref{two-field-lagrangian-simplified-order-4}) which are the
fundaments of studying non-Gaussianity for multi-field models. The
results are as follows\footnote{We here just consider the terms
containing time derivatives and not any spatial derivatives. However
the procedure is same.} for $a^{-3}\Delta{\cal{L}}^{(3)}$
\begin{eqnarray}\label{adi-ent-pert-correction-order-3}\nonumber
&&4\dot\sigma\big(\vec{T}.\dot{\vec\delta}\big)^3\times\bigg[g_1\cos^4\theta+g_2\sin^4\theta+g_3\cos^3\theta\sin\theta
+g_4\cos\theta\sin^3\theta+(g_5+g_6)\cos^2\theta\sin^2\theta\bigg]\\\nonumber&+&
\dot\sigma\big(\vec{N}.\dot{\vec\delta}\big)^3\times\\\nonumber&&\vspace{0cm}
\bigg[-g_4\cos^4\theta+g_3\sin^4\theta-2\bigg(2g_2-(g_5+g_6)\bigg)\cos^3\theta\sin\theta
+2\bigg(2g_1-(g_5+g_6)\bigg)\cos\theta\sin^3\theta+3(g_4-g_3)\cos^2\theta\sin^2\theta\bigg]\\\nonumber&+&
3\dot\sigma\big(\vec{T}.\dot{\vec\delta}\big)^2\big(\vec{N}.\dot{\vec\delta}\big)\times\\\nonumber&&
\bigg[-g_3\cos^4\theta+g_4\sin^4\theta+2\bigg(2g_1-(g_5+g_6)\bigg)\cos^3\theta\sin\theta-2\bigg(2g_2-(g_5+g_6)\bigg)\cos\theta\sin^3\theta
+3(g_3-g_4)\cos^2\theta\sin^2\theta\bigg]\\\nonumber&+&
2\dot\sigma\big(\vec{T}.\dot{\vec\delta}\big)\big(\vec{N}.\dot{\vec\delta}\big)^2\times\\\nonumber&&
\bigg[(g_5+g_6)\bigg(\cos^4\theta+\sin^4\theta\bigg)+3(g_4-g_3)\bigg(\cos^3\theta\sin\theta-\cos\theta\sin^3\theta\bigg)
+\bigg(3(g_1+g_2)-2(g_5+g_6)\bigg)\cos^2\theta\sin^2\theta\bigg]
\end{eqnarray}
and for $a^{-3}\Delta{\cal{L}}^{(4)}$ it becomes
\begin{eqnarray}\label{adi-ent-pert-correction-order-4}\nonumber
&&\big(\vec{T}.\dot{\vec\delta}\big)^4\times\bigg[g_1\cos^4\theta+g_2\sin^4\theta+g_3\cos^3\theta\sin\theta
+g_4\cos\theta\sin^3\theta+(g_5+g_6)\cos^2\theta\sin^2\theta\bigg]\\\nonumber&+&
\big(\vec{N}.\dot{\vec\delta}\big)^4\times
\bigg[g_2\cos^4\theta+g_1\sin^4\theta-g_4\cos^3\theta\sin\theta
-g_3\cos\theta\sin^3\theta+(g_5+g_6)\cos^2\theta\sin^2\theta\bigg]\\\nonumber&+&
\big(\vec{T}.\dot{\vec\delta}\big)^3\big(\vec{N}.\dot{\vec\delta}\big)\times\\\nonumber&&
\bigg[-g_3\cos^4\theta+g_4\sin^4\theta+2\bigg(2g_1-(g_5+g_6)\bigg)\cos^3\theta\sin\theta-2\bigg(2g_2-(g_5+g_6)\bigg)\cos\theta\sin^3\theta
+3(g_3-g_4)\cos^2\theta\sin^2\theta\bigg]\\\nonumber&+&
\big(\vec{T}.\dot{\vec\delta}\big)\big(\vec{N}.\dot{\vec\delta}\big)^3\times\\\nonumber&&
\bigg[-g_4\cos^4\theta+g_3\sin^4\theta-2\bigg(2g_2-(g_5+g_6)\bigg)\cos^3\theta\sin\theta+2\bigg(2g_1-(g_5+g_6)\bigg)\cos\theta\sin^3\theta
+3(g_4-g_3)\cos^2\theta\sin^2\theta\bigg]\\\nonumber&+&
\big(\vec{T}.\dot{\vec\delta}\big)^2\big(\vec{N}.\dot{\vec\delta}\big)^2\times\\\nonumber&&
\bigg[(g_5+g_6)\bigg(\cos^4\theta+\sin^4\theta\bigg)+3(g_4-g_3)\bigg(\cos^3\theta\sin\theta-\cos\theta\sin^3\theta\bigg)
+2\bigg(3(g_1+g_2)-2(g_5+g_6)\bigg)\cos^2\theta\sin^2\theta\bigg].
\end{eqnarray}

Now let us just consider the terms containing
$(\vec{T}.\dot{\vec\delta}\big)$ and write them together
as\footnote{Note that the coefficient of
$\big(\vec{T}.\dot{\vec\delta}\big)^2$ in
(\ref{two-field-lagrangian-simplified-order-2-correction-terms}) is
$6$. But by comparison to
(\ref{two-field-lagrangian-simplified-order-2-correction-terms-spatial}),
$2$ of $6$ appear in definition of the speed of sound, $c_s$,
(exactly same as the second line in (\ref{scalar-perturbation})) and
what remains is $4\big(\vec{T}.\dot{\vec\delta}\big)^2$.}
\begin{eqnarray}\label{Tdelta}\nonumber
4\dot\sigma^2\bigg[g_1\cos^4\theta+g_2\sin^4\theta+g_3\cos^3\theta\sin\theta
+g_4\cos\theta\sin^3\theta+(g_5+g_6)\cos^2\theta\sin^2\theta\bigg]\times
\bigg(\big(\vec{T}.\dot{\vec\delta}\big)^2+
\big(\vec{T}.\dot{\vec\delta}\big)^3/\dot\sigma+\frac{1}{4}\big(\vec{T}.\dot{\vec\delta}\big)^4/\dot\sigma^2\bigg)
\end{eqnarray}
Comparison the above relation with the relation in
(\ref{scalar-perturbation}) manifests that $\sigma$, $[...]$ and
$\big(\vec{T}.\dot{\vec\delta}\big)$ play the role of $\vpb$,
$f(\vpb)$ and $\dot\delvp$ respectively. The significant property of
this model is the appearance of $\big(\vec{T}.\dot{\vec\delta}\big)$
and $\big(\vec{N}.\dot{\vec\delta}\big)$ or equivalently
$(\dot{\delta\sigma}-\dot\theta\delta s)$ and $(\dot{\delta
s}+\dot\theta\delta \sigma)$ respectively. This means that
$\dot{\delta\sigma}$ and $\delta s$ are always together and the same
for $\dot{\delta s}$ and $\delta\sigma$. This characteristic feature
of this model has some observational consequences which are
discussed in the following.

\subsubsection{The Amplitude of Non-Gaussianity}
Now we are going to estimate the non-Gaussianity amplitude. To do
this one procedure is comparison between the non-linear terms and
the linear ones. Mathematically, the amplitude of non-Gaussianity
$f_{NL}$, bi-spectrum, can be estimated as
$\frac{{\cal{L}}^{(3)}}{{\cal{L}}^{(2)}}\times\zeta^{-1}$ where
$\zeta$ is the curvature perturbation \cite{baumann}. Note that the
dominant amplitude of the terms containing time derivatives comes
from their amplitude at horizon crossing. At this time
$\frac{d}{dt}\sim H$ where $H$ is the Hubble constant. Hence for the
second order perturbations the Lagrangian ${\cal{L}}^{(2)}$ can be
written in an abstract form as $\{H^2, H \dot{\theta},
\dot{\theta}^2\}\times M^2 \times \delta\sigma^2$. The same analyze
for ${{\cal{L}}^{(3)}}$ results in $\{H^3, H^2 \dot{\theta}, H
\dot{\theta}^2, \dot{\theta}^3\}\times f(g_i) \times \dot\sigma
\times \delta\sigma^3$. So in an abstract form
\begin{eqnarray}\label{fNL}
\frac{{\cal{L}}^{(3)}}{{\cal{L}}^{(2)}}=\frac{\{H^3, H^2
\dot{\theta}, H \dot{\theta}^2, \dot{\theta}^3\}}{\{H^2, H
\dot{\theta}, \dot{\theta}^2\}}\times\bigg(\frac{f(g_i)}{M^2}\dot\sigma^2\bigg)\times
\frac{\delta\sigma}{\dot\sigma}=\frac{\{H^3, H^2 \dot{\theta}, H
\dot{\theta}^2, \dot{\theta}^3\}}{H\times\{H^2, H \dot{\theta},
\dot{\theta}^2\}}\times\bigg(\frac{f(g_i)}{M^2}\dot\sigma^2\bigg)\times \zeta
\end{eqnarray}
where $\zeta\sim \frac{H\delta\sigma}{\dot\sigma}$ is interpreted as
curvature perturbation. Now we consider two different regimes
$\dot{\theta}<<H$ and $\dot{\theta}>>H$. The first regime,
$\dot{\theta}<<H$, physically means that the model is a single field model effectively. In
this case the amplitude of bi-spectrum can be approximated by
\begin{eqnarray}\label{fNL-thetadot<<H}
f_{NL}\sim\frac{{\cal{L}}^{(3)}}{{\cal{L}}^{(2)}}\zeta^{-1}\sim\frac{f(g_i)}{M^2}\dot\sigma^2
\end{eqnarray}
but remember that the validity of the effective
field theory imposes $\frac{f(g_i)}{M^2}\dot\sigma^2<1$. So in this case as mentioned before
there is no significant non-Gaussianity which is in agreement with
the single field models of inflation \cite{chen}. But the other case,
$\dot{\theta}>>H$, means that the classical path in the
phase space is highly curved \cite{ana}. In other words it means the classical path
in the phase-space is far from a straight line ($\dot\theta=0$).
 So the existence of the entropic field
is unavoidable. For this case the amplitude of $f_{NL}$ is
\begin{eqnarray}\label{fNL-thetadot>>H}
f_{NL}\sim\frac{{\cal{L}}^{(3)}}{{\cal{L}}^{(2)}}\zeta^{-1}
\sim\frac{\dot\theta}{H}\times\bigg(\frac{f(g_i)}{M^2}\dot\sigma^2\bigg).
\end{eqnarray}
Now the factor
$\frac{\dot\theta}{H}\times(\frac{f(g_i)}{M^2}\dot\sigma^2)$ can be
large and results in large non-Gaussianity consequently. So the
large curvature of the classical path in the phase-space results in
the large non-Gaussianity. Though this result can be compared to the
other works in the literature \cite{gpmulti} and in the effective
field theory context \cite{eftmulti} but the curvature of the
classical path is restricted due to observed scale invariant power
spectrum \cite{max}.

For a moment let us relax the constraint on the
$\frac{f(g_i)}{M^2}\dot\sigma^2$. Consequently the correction terms
in (\ref{two-field-lagrangian-simplified}) causes the large
non-Gaussinity. However the relaxation of the constraint can be
justified by assuming that our model is completely described by
(\ref{two-field-lagrangian-simplified}) without any higher order
correction terms. This needs fine tuning which is not impossible but
it is not natural. However there is another method to rationalize
this assumption. Instead of fine tuning the model automatically
shows this property via for example Vainshtein mechanism
\cite{derham,vain}.

The same is applicable for tri-spectrum by an estimation as
$\tau_{NL}\sim\frac{{\cal{L}}^{(4)}}{{\cal{L}}^{(2)}}\times\zeta^{-2}$.
The fourth order Lagrangian, ${{\cal{L}}^{(4)}}$, can be written in
the abstract form as $\{H^4,H^3\dot\theta, H^2 \dot{\theta}^2, H
\dot{\theta}^3,H \dot{\theta}^3,\dot\theta^4\}\times f(g_i)  \times
\delta\sigma^4$ and then
\begin{eqnarray}\label{tNL-thetadot<<H}
\tau_{NL}\sim\frac{{\cal{L}}^{(4)}}{{\cal{L}}^{(2)}}\zeta^{-2}\sim\frac{f(g_i)}{M^2}\dot\sigma^2
\end{eqnarray}
for $\dot\theta<<H$ and
\begin{eqnarray}\label{tNL-thetadot>>H}
\tau_{NL}\sim\frac{{\cal{L}}^{(4)}}{{\cal{L}}^{(2)}}\zeta^{-2}
\sim\left(\frac{\dot\theta}{H}\right)^2\times\bigg(\frac{f(g_i)}{M^2}\dot\sigma^2\bigg).
\end{eqnarray}
for $\dot\theta>>H$.

\subsubsection{The Shape of Non-Gaussianity}
Now let us focus on the shape of possible non-Gaussinity predicted
by our model. In principle all the possible interaction terms
between $\delta\sigma$, $\delta s$ and their derivatives exist in
our model. This fact means all the non-Gaussianity shape can be
produced. However the amplitude of different shapes are different.
As a general argument it can be emphasized that for different
regimes of $\frac{\dot\theta}{H}$ different shapes are dominant. For
the case $\dot\theta<<H$ the terms containing time derivatives
become dominated. This means in this limit the equilateral shape is
the main one among the others. But it does not mean the other shapes
do not exist i.e. the ``Cosine" between different shapes are not
zero. In the other limit, $\dot\theta>>H$, the terms without
derivative become dominant and then the local shape is dominant.
This result is in agreement with the prediction for multi-field
inflation models \cite{chen}. Note that in \cite{senatore} since the
additional entropy perturbations are added by symmetry then they do
not have any term without derivative in their Lagrangian. In this
sense they do not predict a dominant local shape for their model
which is in disagreement with our result.

A characteristic feature of this model is appearance of just two
combinations of the fields i.e. $\vec{T}.\dot{\vec{\delta}}$ and
$\vec{N}.\dot{\vec{\delta}}$ in all the terms including the second,
third and fourth orders. To explain what is the physical result of
this fact let us concentrate on
$\vec{T}.\dot{\vec{\delta}}=\dot{\delta\sigma}-\dot\theta\delta s$,
as an example. The third order term of this combination is
$(\vec{T}.\dot{\vec{\delta}})^3=\dot{\delta\sigma^3}-
3\dot\theta\dot{\delta\sigma^2} \delta
s+3\dot\theta^2\dot{\delta\sigma} \delta s^2 -\dot\theta^3\delta
s^3$. Without worrying about the amplitude in this part let us focus
on the first and the last term. The definite prediction of this
model is that if any equilateral non-Gaussianity due to the first
term, i.e. $\dot{\delta\sigma^3}$, be observed then it has to be
observed a local non-Gaussianity due to the last term\footnote{ Note
that due to the first equation of motion in (\ref{eq-mo-pert}) the
$\delta s$ sources $\delta \sigma$.}, i.e. $\delta s^3$. So the
non-Gaussianity predicted by this model cannot be pure e.g. pure
equilateral shape. Hence mathematically, the ``Cosine" between two
shapes cannot be zero and more the ``Cosine" depends on the
$\dot\theta$ and is fixed by the model. This argument is true for
the other third order terms as well as fourth order ones. To
conclude, it seems that this model predicts a definite combination
of different shapes for the non-Gaussianity if the amplitude allows
to observe them.

\subsection{An Example}
In this subsection we are going to show how the general statements
mentioned before do work in a simple example. Here we assume that
all the $g_i$'s vanish except $g_1(\vp,\chi)$\footnote{Note that
except here in the rest of the paper we assumed that $g_i$'s are
constant as a matter of simplification. But here we would like to
show how the functionality of $g_i$'s may affect the final result.}
which is a generalization of the model in \cite{paolo1}. In addition
we assume there is no potential term\footnote{Note that the most
general form of the potential term can be supposed. But as mentioned
in \cite{baumann-green}, in the slow-roll regime there is no
interesting non-Gaussianity prediction for single field models.
However for multi-field models the potential term can result in
large non-Gaussianity which considered in \cite{gpmulti}. Here, we
restrict our calculations to kinetic terms.}. According to the
background Lagrangian
(\ref{two-field-lagrangian-simplified-order-0}) the equations of
motion for our special case become
\begin{eqnarray}\label{special-case-eq.of.motion}
&&\ddot{\vp}\big(1+12
\frac{g_1}{M^2}\dot\vp^2\big)+3H\dot\vp\big(1+4
\frac{g_1}{M^2}\dot\vp^2\big)+\frac{1}{M^2}\dot\vp^3\bigg(3\dot\vp\frac{\partial
g_1}{\partial \vp}+4\dot\chi\frac{\partial g_1}{\partial
\chi}\bigg)=0\\\nonumber
&&\ddot{\chi}+3H\dot\chi-\frac{1}{M^2}\dot\vp^4\frac{\partial
g_1}{\partial\chi}=0
\end{eqnarray}
where $M=M_1=M_2$ is assumed. On the other hand, what can cause the
significant non-Gaussianity is $\dot\theta$ as mentioned before. In
general due to the definition of $\theta$ in
(\ref{tangent-normal-vector}), $\dot\theta$ can be read as
\begin{eqnarray}\label{dot-theta}\nonumber
\dot\theta=\frac{-\ddot\vp\dot\chi+\dot\vp\ddot\chi}{\dot\vp^2+\dot\chi^2}
\end{eqnarray}
and in our special case by considering
(\ref{special-case-eq.of.motion}) it becomes (up to the first order
of $g_1\frac{\dot\vp^2}{M^2}$)
\begin{eqnarray}\label{dot-theta-special-case}\nonumber
\frac{\dot\theta}{H}\simeq24 \bigg(-g_1
\frac{\dot\vp^2}{M^2}\bigg)\frac{\dot\vp\dot\chi}{\dot\vp^2+\dot\chi^2}+\frac{1}{H}
\frac{1}{M^2}\frac{\dot\vp^3}{\dot\vp^2+\dot\chi^2}\bigg[3\frac{\partial
g_1}{\partial\vp}\dot\vp\dot\chi+\frac{\partial
g_1}{\partial\chi}\big(\dot\vp^2+4\dot\chi^2\big)\bigg].
\end{eqnarray}
The condition $g_1 \frac{\dot\vp^2}{M_1^2}<1$ ensures the validity
of the effective field theory. So it is not bad to estimate $24g_1
\frac{\dot\vp^2}{M_1^2}\sim 1$. Then due to the first term
$\frac{\dot\theta}{H}\sim
\frac{\dot\vp\dot\chi}{\dot\vp^2+\dot\chi^2}$ which means the
maximum of $f_{NL}$ in (\ref{fNL-thetadot>>H}) is less than one. A
successful inflation in the slow-roll regime restricts the value of
the field velocities which may restrict more the above estimation.
To discuss on the second term let us divide $g_1(\vp,\chi)$ to its
amplitude and functionality as $g_1(\vp,\chi)=\mid g_1\mid\times
f(\vp,\chi)$ such that $\mid g\mid$ is the amplitude of the
$g_1(\vp,\chi)$ and $f(\vp,\chi)$ represents its functional form. So
the second term can be estimated as (by assuming
$\dot\vp\sim\dot\chi$)
\begin{eqnarray}\label{dot-theta-special-case-second-term}\nonumber
\frac{\dot\theta}{H}\simeq  \bigg(\mid
g_1\mid\frac{\dot\vp^2}{M^2}\bigg)\frac{\dot\vp}{2H}\bigg[3\frac{\partial
f}{\partial\vp}+5\frac{\partial f}{\partial\chi}\bigg],
\end{eqnarray}
where $\mid g_1\mid \frac{\dot\vp^2}{M^2}<1$ has to be satisfied. On
the other hand one of the Friedmann equations (in the absence of the
potential) is
$H^2=\frac{M^2}{2}\dot\vp^2+\frac{M^2}{2}\dot\chi^2\sim
M^2\dot\vp^2$ for the zeroth order of $g_1 \frac{\dot\vp^2}{M^2}$.
Now if $\frac{\partial f}{\partial\vp}$ or
$\frac{\partial f}{\partial\chi}$ have the significant
amplitude with respect to $M$ then a large amplitude of
non-Gaussianity would be expected. This can be realized by
assuming sharp features in the functionality of $g_1(\vp,\chi)$
maybe due to a phase transition.

\subsection{Some Clarifications on Differences with Senatore and
Zaldarriaga \cite{senatore}} The significant difference is the
existence of the terms containing the adiabatic and entropy
perturbations themselves not just their derivatives. The reason for
this difference is in how the effective field theory is constructed
in \cite{senatore}. As mentioned before in their model the adiabatic
mode is borrowed from \cite{paolo} which satisfies a shift symmetry.
Then the entropy modes are added and satisfy the shift symmetry too.
Consequently, in their formalism they have just derivative of
perturbations. But in contrast we do not start with distinguishable
fields then we do not have any difference between the perturbations
initially. So by this starting point we had to define the adiabatic
and entropy perturbations. This is what has been done in this
section in details. Now the question is that is there any special
transformation for adiabatic and entropy perturbation in our model?
Yes, it is locally rotated shift transformation i.e.
\begin{eqnarray}\label{locally-rotated-shift-symm.}
\delta\sigma\rightarrow\delta\sigma+(c_1
\cos\theta+c_2\sin\theta)\\\nonumber \delta s\rightarrow\delta
s+(c_1 \sin\theta-c_2\cos\theta)
\end{eqnarray}
where $\theta=\arctan(\dot\chi/\dot\vp)$. Note that the rotational
angle depends on the background fields time evolution. To achieve
this result, the starting point is the Lagrangian for two fields
i.e. the relations (\ref{two-field-lagrangian-simplified-order-2}),
(\ref{two-field-lagrangian-simplified-order-3}) and
(\ref{two-field-lagrangian-simplified-order-4}). By looking at these
relations it is obvious that $\delvp\rightarrow\delvp+c_1$ and
$\delchi\rightarrow\delchi+c_2$ is a symmetry of the model where
$c_1$ and $c_2$ are two independent arbitrary constants. So due to
(\ref{adi-ent-perturbations}) one can get the above relation
(\ref{locally-rotated-shift-symm.}) as the corresponding
transformation of $\delta \sigma$ and $\delta s$. According to this
relation the invariant terms corresponding to $\dot{\delvp}$ and
$\dot{\delta \chi}$ are not $\dot{\delta\sigma}$ and $\dot{\delta
s}$ but
\begin{eqnarray}\label{invariant-combination}
\dot{\delta\sigma}-\dot\theta\delta
s\rightarrow\dot{\delta\sigma}-\dot\theta\delta
s\\\nonumber\dot{\delta s}+\dot\theta\delta
\sigma\rightarrow\dot{\delta s}+\dot\theta\delta \sigma
\end{eqnarray}
which are $\vec T.\dot{\vec\delta}$ and $\vec N.\dot{\vec\delta}$
respectively. Not surprisingly, these terms construct whole
Lagrangian in adiabatic and entropy perturbations language
as seen previously.

So it seems initially supposed adiabatic perturbation in
\cite{senatore} results in lack of all possible terms in the
effective Lagrangian. Our proposition to solve this problem is based
on the discussion in this subsection. The main building blocks for
an effective field theory of multi-field inflation are not
$\dot{\delta\sigma}$ and $\dot{\delta s}$ but they are  $\vec
T.\dot{\vec\delta}$ and $\vec N.\dot{\vec\delta}$. So the most
general Lagrangian for the perturbations in the multi-field context
should be written as\footnote{Note that here we just consider the
time derivative since in the discussion of this section there is no
difference between our model and \cite{senatore} for the terms
containing spatial derivatives. This is because the background is
not spatial dependent. Remember that the angle of rotation just
depends on time.}
\begin{eqnarray}\label{general-perturbation-lag.-adi-ent.}
\Delta{\cal{L}}\propto \sum_{m,n} c_{mn} \bigg(\vec
T.\dot{\vec\delta}\bigg)^m\bigg(\vec N.\dot{\vec\delta}\bigg)^n
\end{eqnarray}
for arbitrary $c_{mn}$. The above Lagrangian can be considered as
the effective field theory for the two-field inflation in the
language of \cite{senatore} but with additional terms. Note that the
above result can be generalized to multi-field inflation as
\begin{eqnarray}\label{general-perturbation-lag.-adi-ent.-multi}
\Delta{\cal{L}}\propto \sum c_{{n_0},{n_1},...,{n_N}} \bigg(\vec
T.\dot{\vec\delta}\bigg)^{n_0}\bigg(\vec
N_1.\dot{\vec\delta}\bigg)^{n_1}\bigg(\vec
N_2.\dot{\vec\delta}\bigg)^{n_2}...\bigg(\vec
N_N.\dot{\vec\delta}\bigg)^{n_N}
\end{eqnarray}
where $\vec T$ and $\vec N_i$'s are a set of orthonormal vectors for
an $(N+1)$-field model.

\section{Conclusions}
In this work the effective field theory of multi-field inflation has
been studied as a generalization of Weinberg's idea \cite{weinberg}
for a single field. In this approach the most general Lagrangian is
built by using all the covariant terms of the fields. Though
effectively the terms with higher order derivatives are interested
in the higher energy scales. In this work we restricted the model to
the first correction terms. They results in up to fourth order terms
in perturbations. Then due to the physical interests we switched to
the adiabatic and entropy formalism. It has been shown that
generally these modes can have different speeds of sound. By
considering the non-linear terms we studied the non-Gaussinity in
this model. It has been shown that the amplitude of non-Gaussianity
can be significant when the curvature of the classical path in the
phase-space becomes large. For example a sharp turn in the classical
path can realize it. However it seems that existence of the higher
order derivative terms in the Lagrangian cannot produce large
non-Gaussinity. The bottom line for this fact is the strong
constraint on the coefficients to keep the effective field theory
valid. But there is an idea that it is possible to take the higher
order correction terms under control automatically, e.g. by
Vainshtein mechanism. This relaxes the constraint on the coefficient
of the correction terms and results in large non-Gaussianity. On the
other hand the structure of the interacting terms in the Lagrangian
predicts the existence of all the shapes of non-Gaussinity with the
different amplitude for different cases. But the characteristic
feature of the model is that the non-Gaussinities are correlated.
That means if there is a local non-Gaussinity due to the entropy
mode then certainly there is a non-Gaussinity in adiabatic mode
which is equilateral. The amplitude of these different types of
non-Gaussinities are not independent.

In contrast to \cite{senatore}, the adiabatic and entropy
perturbations are not distinguishable initially. This fact results
in the existence of the perturbations as well as their derivatives.
In other words the adiabatic and entropy perturbations are not
invariant under the shift symmetry of original fields. However a
combination of them is invariant under such a symmetry. These
combinations are ``$\dot{\delta\sigma}-\dot\theta\delta s$" and
``$\dot{\delta s}+\dot\theta\delta \sigma$" or in other form $\vec
T.\dot{\vec\delta}$ and $\vec N.\dot{\vec\delta}$ respectively. This
result is important for constructing the effective field theory for
multi-field inflation and causes the additional terms with respect
to what is considered in \cite{senatore}.
%===============================================================================
%=============================================================
%\vspace{1cm}
\begin{acknowledgments}
We would like to thank B. A. Bassett, H. Firouzjahi, J. Fonseca, H.
R. Sepnagi, N. Sivanandam and M. M. Sheikh-Jabbari for their
comments. We are grateful of T. Battefeld for his very useful
comments and careful reading of the manuscript. We also specially
thank P. Creminelli for very fruitful discussions and comments. We
would like to thank ICTP for their warm hospitality and support when
this work was initiated.
\end{acknowledgments}
\appendix
%{\textbf{Appendix A: The Most General Lagrangian for Multi-Field
%Inflation}}\label{appendixA}
\section{The Most General Lagrangian for Multi-Field
Inflation}\label{appendixA} In this appendix we will study the most
general form of a Lagrangian including up to the fourth order
space-time derivatives for multi-field models. It is a
straightforward generalization of what has been done in
\cite{elizalde} for a single field. To continue let us review the
blocks
\begin{eqnarray}\label{blocks}\nonumber
\vp,\nabla_\mu\vp,\nabla_\mu\nabla_\nu\vp,\nabla_\mu\nabla_\nu\nabla_\rho\vp,\nabla_\mu\nabla_\nu\nabla_\rho\nabla_\sigma\vp,
g_{\mu\nu},R_{\mu\nu\rho\sigma},\nabla_{\alpha}R_{\mu\nu\rho\sigma},\nabla_{\alpha}\nabla_{\beta}R_{\mu\nu\rho\sigma}
\end{eqnarray}
which satisfy general covariance. Obviously all the indices can be
raised by the metric inverse $g^{\mu\nu}$. To construct the
Lagrangian we need to make scalars from the above blocks. The
possible terms are listed in \cite{elizalde} and easily can be
generalized for more than one field. The most general form of the
Lagrangian with the above blocks for a multi field model reads as
${\cal{L}}/\sqrt{g}=$
\begin{eqnarray}\label{most-general-lagrangian}
&\bigg\{&b_1^{IJ}(\vec\vp)\Box\vp_I\Box\vp_J+b_2^{IJK}(\vec\vp)\nabla_\mu\vp_I\nabla^\mu\vp_J\Box\vp_K+
b_3^{IJKL}(\vec\vp)\nabla_\mu\vp_I\nabla^\mu\vp_J\nabla_\nu\vp_K\nabla^\nu\vp_L+b_4^{IJ}(\vec\vp)\nabla_\mu\vp_I\nabla^\mu\vp_J\\\nonumber
&+&b_5(\vec\vp)+b^{IJ}_6(\vec\vp)(\nabla^\mu\vp_I)(\Box\nabla_\mu\vp_J)+b^I_7(\vec\vp)(\nabla_\mu\nabla_\nu\vp_I)^2+b^{IJK}_8(\vec\vp)(\nabla_\mu\vp_I)
(\nabla_\nu\vp_J)(\nabla^\mu\nabla^\nu\vp_K)+b^I_9(\vec\vp)\nabla^\mu\Box\nabla_\mu\vp_I\\\nonumber&+&b^I_{10}(\vec\vp)\Box^2\vp_I
+b^I_{11}(\vec\vp)(\nabla^\mu)\nabla_\mu\Box\vp_I
+c_1^{IJ}(\vec\vp)R\nabla_\mu\vp_I\nabla^\mu\vp_J+c_2^{IJ}(\vec\vp)R^{\mu\nu}\nabla_\mu\vp_I\nabla_\nu\vp_J+c_3^{I}(\vec\vp)R\Box\vp_I\\\nonumber
&+&c^I_4(\vec\vp)(\nabla^\mu R)(\nabla_\mu\vp_I)+c_5(\vec\vp)\Box
R+c^I_6(\vec\vp)
R_{\mu\nu}\nabla^\mu\nabla^\nu\vp_I+a_1(\vec\vp)R_{\mu\nu\rho\sigma}R^{\mu\nu\rho\sigma}+a_2(\vec\vp)R_{\mu\nu}R^{\mu\nu}+a_3(\vec\vp)R^2+a_4(\vec\vp)R\bigg\}
\end{eqnarray}
where summation on repeated indices is assumed and
$\vec\vp=\vp_1,\vp_2,...,\vp_N$ for $N$ fields. Note that all of the
terms in (\ref{most-general-lagrangian}) are not independent up to
total derivative terms. In the following we show this fact in
details.
\begin{eqnarray}\label{total-derivative-term}\nonumber
c^I_4(\vec\vp)(\nabla^\mu
R)(\nabla_\mu\vp_I)&=&\nabla^\mu\big(c^I_4(\vec\vp)R\nabla_\mu\vp_I\big)-c^I_4(\vec\vp)R\Box\vp_I-R\big(\nabla^\mu
c^I_4(\vec\vp)\big)\nabla_\mu\vp_I\\\nonumber
&=&\nabla^\mu\big(c^I_4(\vec\vp)R\nabla_\mu\vp_I\big)-c^I_4(\vec\vp)R\Box\vp_I-R\big(\partial
c^I_4(\vec\vp)/\partial\vp_J\big)\nabla^\mu\vp_J
\nabla_\mu\vp_I
\end{eqnarray}
\begin{eqnarray}\label{total-derivative-term1}\nonumber
c_5(\vec\vp)\Box
R&=&\nabla^\mu(c_5(\vec\vp)\nabla_\mu R)-(\nabla^\mu
c_5(\vec\vp))(\nabla_\mu
R)\\\nonumber&=&\nabla^\mu\bigg((c_5(\vec\vp)\nabla_\mu
R)-((\nabla_\mu c_5(\vec\vp)) R)\bigg)+(\nabla_\mu\nabla^\mu
c_5(\vec\vp))R\\\nonumber&=&\nabla^\mu\bigg((c_5(\vec\vp)\nabla_\mu
R)-((\nabla_\mu c_5(\vec\vp)) R)\bigg)+\nabla_\mu\bigg((\partial
c_5(\vec\vp)/\partial\vp_I)\nabla^\mu\vp_I\bigg)R\\\nonumber&=&\nabla^\mu\bigg((c_5(\vec\vp)\nabla_\mu
R)-((\nabla_\mu c_5(\vec\vp)) R)\bigg)+\bigg(\frac{\partial^2
c_5(\vec\vp)}{\partial\vp_I\partial\vp_J}\nabla_\mu\vp_I\nabla^\mu\vp_J+(\partial
c_5(\vec\vp)/\partial\vp_I)\nabla_\mu\nabla^\mu\vp_I\bigg)R
\end{eqnarray}
\begin{eqnarray}\label{total-derivative-term2}\nonumber
c_6^I(\vec\vp)R_{\mu\nu}(\nabla^\mu\nabla^\nu\vp_I)&=&\nabla^\mu\bigg(c_6^I(\vec\vp)R_{\mu\nu}\nabla^\nu\vp_I\bigg)
-R_{\mu\nu}\big(\nabla^\mu
c_6^I(\vec\vp)\big)\nabla^\nu\vp_I-c_6^I(\vec\vp)\big(\nabla^\mu
R_{\mu\nu}\big)\nabla^\nu\vp_I\\\nonumber
&=&\nabla^\mu\bigg(c_6^I(\vec\vp)R_{\mu\nu}\nabla^\nu\vp_I\bigg)
-R_{\mu\nu}\big(\partial
c_6^I(\vec\vp)/\partial\vp_J\big)\nabla^\mu\vp_J\nabla^\nu\vp_I+\frac{1}{2}g_{\mu\nu}c_6^I(\vec\vp)\big(\nabla^\mu
R\big)\nabla^\nu\vp_I
\\\nonumber
&=&\nabla^\mu\bigg(c_6^I(\vec\vp)R_{\mu\nu}\nabla^\nu\vp_I\bigg)
-R_{\mu\nu}\big(\partial
c_6^I(\vec\vp)/\partial\vp_J\big)\nabla^\mu\vp_J\nabla^\nu\vp_I+\frac{1}{2}g_{\mu\nu}\nabla^\mu\bigg(c_6^I(\vec\vp)
R\nabla^\nu\vp_I\bigg)\\\nonumber&-&\frac{1}{2}g_{\mu\nu}R\big(\nabla^\mu
c_6^I(\vec\vp)
\big)\nabla^\nu\vp_I-\frac{1}{2}g_{\mu\nu}c_6^I(\vec\vp)
R\nabla^\mu\nabla^\nu\vp_I\\\nonumber&=&\nabla^\mu\bigg(c_6^I(\vec\vp)R_{\mu\nu}\nabla^\nu\vp_I\bigg)
+\frac{1}{2}\nabla^\mu\bigg(c_6^I(\vec\vp)
R\nabla_\mu\vp_I\bigg)\\\nonumber&-&R_{\mu\nu}\big(\partial
c_6^I(\vec\vp)/\partial\vp_J\big)\nabla^\mu\vp_J\nabla^\nu\vp_I-\frac{1}{2}R\big(\partial
c_6^I(\vec\vp)/\partial\vp_J\big)\nabla_\mu\vp_J\nabla^\mu\vp_I-\frac{1}{2}g_{\mu\nu}c_6^I(\vec\vp)
R\nabla^\mu\nabla^\nu\vp_I
\end{eqnarray}
\begin{eqnarray}\label{total-derivative-term3}\nonumber
b_6^{IJ}(\vec\vp)(\nabla^\mu\vp_I)\Box\nabla_\mu\vp_J&=&b_6^{IJ}(\vec\vp)(\nabla^\mu\vp_I)\nabla^\nu\nabla_\nu\nabla_\mu\vp_J=
b_6^{IJ}(\vec\vp)(\nabla^\mu\vp_I)\nabla^\nu\nabla_\mu\nabla_\nu\vp_J+b_6^{IJ}(\vec\vp)(\nabla^\mu\vp_I)\nabla^\nu[\nabla_\nu,\nabla_\mu]\vp_J\\\nonumber
&=&b_6^{IJ}(\vec\vp)(\nabla^\mu\vp_I)\nabla^\nu\nabla_\mu\nabla_\nu\vp_J+b_6^{IJ}(\vec\vp)(\nabla^\mu\vp_I)\nabla^\nu[\nabla_\nu,\nabla_\mu]\vp_J\\\nonumber
&=&b_6^{IJ}(\vec\vp)(\nabla^\mu\vp_I)\nabla_\mu\Box\vp_J+b_6^{IJ}(\vec\vp)(\nabla^\mu\vp_I)[\nabla_\nu,\nabla_\mu]\nabla^\nu\vp_J+
b_6^{IJ}(\vec\vp)(\nabla^\mu\vp_I)\nabla^\nu[\nabla_\nu,\nabla_\mu]\vp_J\\\nonumber
&=&b_6^{IJ}(\vec\vp)(\nabla^\mu\vp_I)\nabla_\mu\Box\vp_J+b_6^{IJ}(\vec\vp)(\nabla^\mu\vp_I)R_{\mu\nu}\nabla^\nu\vp_J+
0\\\nonumber
&=&\nabla_\mu\bigg(b_6^{IJ}(\vec\vp)(\nabla^\mu\vp_I)\Box\vp_J\bigg)-(\nabla_\mu
b_6^{IJ}(\vec\vp))(\nabla^\mu\vp_I)\Box\vp_J-b_6^{IJ}(\vec\vp)(\Box\vp_I)(\Box\vp_J)
\\\nonumber&+&b_6^{IJ}(\vec\vp)R_{\mu\nu}(\nabla^\mu\vp_I)(\nabla^\nu\vp_J)
\\\nonumber
&=&\nabla_\mu\bigg(b_6^{IJ}(\vec\vp)(\nabla^\mu\vp_I)\Box\vp_J\bigg)-(\partial
b_6^{IJ}(\vec\vp)/\partial\vp_K)(\nabla_\mu\vp_K)(\nabla^\mu\vp_I)\Box\vp_J\\\nonumber&-&b_6^{IJ}(\vec\vp)(\Box\vp_I)(\Box\vp_J)
+b_6^{IJ}(\vec\vp)R_{\mu\nu}(\nabla^\mu\vp_I)(\nabla^\nu\vp_J)
\end{eqnarray}
\begin{eqnarray}\label{total-derivative-term4}\nonumber
b_7^{IJ}(\vec\vp)(\nabla_\mu\nabla_\nu\vp_I)(\nabla^\mu\nabla^\nu\vp_J)&=&\nabla_\mu\bigg(b_7^{IJ}(\vec\vp)(\nabla_\nu\vp_I)
(\nabla^\mu\nabla^\nu\vp_J)\bigg)-(\nabla_\mu
b_7^{IJ}(\vec\vp))(\nabla_\nu\vp_I)
(\nabla^\mu\nabla^\nu\vp_J)-b_7^{IJ}(\vec\vp)(\nabla^\nu\vp_I)
(\Box\nabla_\nu\vp_J)\\\nonumber
&=&\nabla_\mu\bigg(b_7^{IJ}(\vec\vp)(\nabla_\nu\vp_I)
(\nabla^\mu\nabla^\nu\vp_J)\bigg)-(\partial
b_7^{IJ}(\vec\vp)/\partial\vp_K)(\nabla_\mu\vp_K)(\nabla_\nu\vp_I)
(\nabla^\mu\nabla^\nu\vp_J)\\\nonumber&-&b_7^{IJ}(\vec\vp)(\nabla^\nu\vp_I)
(\Box\nabla_\nu\vp_J)
\end{eqnarray}
\begin{eqnarray}\label{total-derivative-term5}\nonumber
\nonumber
b_8^{IJK}(\vec\vp)(\nabla_\nu\vp_I)(\nabla_\mu\vp_J)(\nabla^\nu\nabla^\mu\vp_K)&=&
\nabla^\nu\bigg(b_8^{IJK}(\vec\vp)(\nabla_\nu\vp_I)(\nabla_\mu\vp_J)(\nabla^\mu\vp_K)\bigg)b_8^{IJK}(\vec\vp)(\Box\vp_I)(\nabla_\mu\vp_J)(\nabla^\mu\vp_K)
\\\nonumber&-&
b_8^{IJK}(\vec\vp)(\nabla_\nu\vp_I)(\nabla^\nu\nabla_\mu\vp_J)(\nabla^\mu\vp_K)-
(\nabla^\nu(b_8^{IJK}(\vec\vp))(\nabla_\nu\vp_I)(\nabla_\mu\vp_J)(\nabla^\mu\vp_K)\\\nonumber
&=&
\nabla^\nu\bigg(b_8^{IJK}(\vec\vp)(\nabla_\nu\vp_I)(\nabla_\mu\vp_J)(\nabla^\mu\vp_K)\bigg)-b_8^{IJK}(\vec\vp)(\Box\vp_I)(\nabla_\mu\vp_J)(\nabla^\mu\vp_K)\\\nonumber&-&
b_8^{IJK}(\vec\vp)(\nabla_\nu\vp_I)(\nabla^\nu\nabla_\mu\vp_J)(\nabla^\mu\vp_K)-
(\partial
b_8^{IJK}(\vec\vp)/\partial\vp_L)(\nabla^\nu\vp_L)(\nabla_\nu\vp_I)(\nabla_\mu\vp_J)(\nabla^\mu\vp_K)
\end{eqnarray}
\begin{eqnarray}\label{total-derivative-term8}
\nonumber
b_9^I(\vec\vp)\nabla^\nu\Box\nabla_\nu\vp_I&=&\nabla^\nu\bigg(b_9^I(\vec\vp)\Box\nabla_\nu\vp_I\bigg)-(\nabla^\nu
b_9^I(\vec\vp))\Box\nabla_\nu\vp_I=\nabla^\nu\bigg(b_9^I(\vec\vp)\Box\nabla_\nu\vp_I\bigg)-(\partial
b_9^I(\vec\vp)/\partial\vp_J)(\nabla^\nu\vp_J)(\Box\nabla_\nu\vp_I)
\end{eqnarray}
\begin{eqnarray}\label{total-derivative-term6}\nonumber
b_{10}^I(\vec\vp)\Box^2\vp_I&=&\nabla^\mu\bigg(b_{10}^I(\vec\vp)\nabla_\mu\Box\vp_I\bigg)-(\nabla^\mu
b_{10}^I(\vec\vp))\nabla_\mu\Box\vp_I\\\nonumber
&=&\nabla^\mu\bigg(b_{10}^I(\vec\vp)\nabla_\mu\Box\vp_I\bigg)-(\partial
b_{10}^I(\vec\vp)/\partial\vp_J)(\nabla^\mu\vp_J)\nabla_\mu\Box\vp_I\\\nonumber
&=&\nabla^\mu\bigg(b_{10}^I(\vec\vp)\nabla_\mu\Box\vp_I\bigg)-\nabla_\mu\bigg((\partial
b_{10}^I(\vec\vp)/\partial\vp_J)(\nabla^\mu\vp_J)\Box\vp_I\bigg)\\\nonumber&+&\big(\nabla_\mu(\partial
b_{10}^I(\vec\vp)/\partial\vp_J)\big)(\nabla^\mu\vp_J)\Box\vp_I+(\partial
b_{10}^I(\vec\vp)/\partial\vp_J)(\Box\vp_J)\Box\vp_I\\\nonumber
&=&\nabla^\mu\bigg(b_{10}^I(\vec\vp)\nabla_\mu\Box\vp_I\bigg)-\nabla_\mu\bigg((\partial
b_{10}^I(\vec\vp)/\partial\vp_J)(\nabla^\mu\vp_J)\Box\vp_I\bigg)\\\nonumber&+&\bigg(\frac{\partial^2
b_{10}^I(\vec\vp)}{\partial\vp_J\partial\vp_K}\bigg)(\nabla_\mu\vp_K)(\nabla^\mu\vp_J)\Box\vp_I+(\partial
b_{10}^I(\vec\vp)/\partial\vp_J)(\Box\vp_J)\Box\vp_I
\end{eqnarray}
\begin{eqnarray}\label{total-derivative-term7}\nonumber
b_{11}^{IJ}(\vec\vp)(\nabla^\nu\vp_I)(\nabla_\nu\Box\vp_J)&=&\nabla^\nu\bigg(\vp_Ib_{11}^{IJ}(\vec\vp)(\nabla_\nu\Box\vp_J)\bigg)-
\vp_I(\nabla^\mu b_{11}^{IJ}(\vec\vp))(\nabla_\nu\Box\vp_J)-\vp_I
b_{11}^{IJ}(\vec\vp)\Box^2\vp_J\\\nonumber
&=&\nabla^\nu\bigg(\vp_Ib_{11}^{IJ}(\vec\vp)(\nabla_\nu\Box\vp_J)\bigg)-
\bigg(\vp_I(\partial
b_{11}^{IJ}(\vec\vp)/\partial\vp_K)\bigg)\nabla^\nu\vp_K(\nabla_\nu\Box\vp_J)-\vp_I
b_{11}^{IJ}(\vec\vp)\Box^2\vp_J
\end{eqnarray}
Due to these relations, the most general Lagrangian
(\ref{most-general-lagrangian}) reduces to a Lagrangian with
redefined coefficients and vanishing $b_6$, $b_7$, $b_8$, $b_9$,
$b_{10}$, $b_{11}$, $c_4$, $c_5$ and $c_6$.

In this step let us deduce the equations of motion for the leading
terms. To do this let us assume $a_4(\vec\vp)=-\frac{M_P^2}{2}$,
$b_4^{IJ}(\vec\vp)=-\frac{M^2}{2}\delta^{IJ}$ and
$b_5(\vec\vp)=-M_P^2U(\vec\vp)$ by redefinition of the fields
without any loss of generality. Now the equations of motion for the
$\vp_I$ and $g_{\mu\nu}$ are
\begin{eqnarray}\label{eq.motion-L0}
M^2 \Box \vp_I=M_P^2\partial U(\vec\vp)/\partial
\vp_I,\hspace{2cm}R_{\mu\nu}=-(M^2/M_P^2)\delta^{IJ}\nabla_\mu\vp_{I}\nabla_\nu\vp_{J}-U(\vec\vp)g_{\mu\nu}.
\end{eqnarray}
Now by plugging the above relations to the remaining terms of the
Lagrangian (\ref{most-general-lagrangian}), it is easy to show that
\begin{eqnarray}\nonumber
a_3(\vec\vp)R^2&=&a_3(\vec\vp)\bigg(-R_{\mu\nu\rho\sigma}R^{\mu\nu\rho\sigma}+4R_{\mu\nu}
R^{\mu\nu}\bigg)\longrightarrow
a_1(\vec\vp)R_{\mu\nu\rho\sigma}R^{\mu\nu\rho\sigma}+a_2(\vec\vp)R_{\mu\nu}R^{\mu\nu}.
\end{eqnarray}
\begin{eqnarray}\label{simplification}\nonumber
b_1^{IJ}(\vec\vp)\Box\vp_I\Box\vp_J&=&\frac{M_P^4}{M^4}b_1^{IJ}(\vec\vp)\frac{\partial
U(\vec\vp)}{\partial \vp_I}\frac{\partial U(\vec\vp)}{\partial
\vp_J} \longrightarrow U(\vec\vp)\end{eqnarray}
\begin{eqnarray}\nonumber
b_2^{IJK}(\vec\vp)\nabla_\mu\vp_I\nabla^\mu\vp_J\Box\vp_K&=&\frac{M_P^2}{M^2}\bigg(b_2^{IJK}(\vec\vp)\frac{\partial
U(\vec\vp)}{\partial
\vp_K}\bigg)\nabla_\mu\vp_I\nabla^\mu\vp_J\longrightarrow
b_4^{IJ}(\vec\vp)\nabla_\mu\vp_I\nabla^\mu\vp_J\end{eqnarray}
\begin{eqnarray}\nonumber
c_1^{IJ}(\vec\vp)R\nabla_\mu\vp_I\nabla^\mu\vp_J&=&c_1^{IJ}(\vec\vp)\bigg(-\frac{M^2}{M_P^2}\delta^{KL}\nabla_\nu\vp_{K}
\nabla^\nu\vp_{L}-4U(\vec\vp)\bigg)\nabla_\mu\vp_I\nabla^\mu\vp_J\\\nonumber&&\longrightarrow
b_3^{IJKL}(\vec\vp)\nabla_\mu\vp_I\nabla^\mu\vp_J\nabla_\nu\vp_K\nabla^\nu\vp_L+b_4^{IJ}(\vec\vp)\nabla_\mu\vp_I\nabla^\mu\vp_J\end{eqnarray}
\begin{eqnarray}\nonumber
c_2^{IJ}(\vec\vp)R^{\mu\nu}\nabla_\mu\vp_I\nabla_\nu\vp_J&=&c_2^{IJ}(\vec\vp)
\bigg(-\frac{M^2}{M_P^2}\delta^{KL}\nabla^\mu\vp_{K}\nabla^\nu\vp_{L}-U(\vec\vp)g^{\mu\nu}\bigg)\nabla_\mu\vp_I\nabla_\nu\vp_J\\\nonumber
&&\longrightarrow
b_3^{IJKL}(\vec\vp)\nabla_\mu\vp_I\nabla^\mu\vp_J\nabla_\nu\vp_K\nabla^\nu\vp_L+b_4^{IJ}(\vec\vp)\nabla_\mu\vp_I\nabla^\mu\vp_J\end{eqnarray}
\begin{eqnarray}\nonumber
c_3^{I}(\vec\vp)R\Box\vp_I&=&\frac{M_P^2}{M^2}c_3^{I}(\vec\vp)\frac{\partial
U(\vec\vp)}{\partial \vp_I}R\longrightarrow
a_4(\vec\vp)R\end{eqnarray} Due to the above relations and again a
redefinition of $\vec\vp$, $U(\vec\vp)$, $b_3^{IJKL}(\vec\vp)$,
$a_1(\vec\vp)$ and $a_2(\vec\vp)$ the most general form of the
Lagrangian becomes
\begin{eqnarray}\label{most-general-lagrangian-simplified-appendix}
{\cal{L}}&=&\sqrt{g}\bigg\{b_3^{IJKL}(\vec\vp)\nabla_\mu\vp_I\nabla^\mu\vp_J\nabla_\nu\vp_K\nabla^\nu\vp_L-\frac{M^2}{2}\delta^{IJ}\nabla_\mu\vp_I\nabla^\mu\vp_J
-M_P^2U(\vec\vp)\\\nonumber
&+&a_1(\vec\vp)R_{\mu\nu\rho\sigma}R^{\mu\nu\rho\sigma}+a_2(\vec\vp)R_{\mu\nu}R^{\mu\nu}-\frac{M_P^2}{2}R\bigg\}
\end{eqnarray}
which exactly reduces to what is deduced in \cite{weinberg} for a
single field model.
\section{Comparison With Gordon et al.
\cite{gordon}}\label{appendixB} It is obvious that the LHS of
(\ref{eq-mo-pert}) is exactly compatible with equations (47) and
(48) in \cite{gordon}. The RHS of equation (47) in \cite{gordon} can
be written as $2\dot\theta \dot{\delta
s}+2(\ddot\theta-\frac{V_{\sigma}}{\dot\sigma}\dot\theta)\delta s$.
So to show that our calculation is compatible with this result one
has to show that
$\ddot\theta+(3H+2\frac{\ddot\sigma}{\dot\sigma})\dot\theta+V_{\sigma
s}=0$. To show the RHS is also the same let us start by pointing out
that
\begin{eqnarray}\label{theta}
&&\dot\theta=\frac{1}{\dot\sigma^2}(\dot\vp\ddot\chi-\dot\chi\ddot\vp)=\frac{1}{\dot\sigma^2}(\dot\chi
V_\vp-\dot\vp V_\chi)\hspace{.5cm}\longrightarrow \hspace{.5
cm}\ddot\theta=-2\frac{\ddot\sigma}{\dot\sigma^3}(\dot\chi
V_\vp-\dot\vp V_\chi)+\frac{1}{\dot\sigma^2}(\ddot\chi
V_\vp-\ddot\vp V_\chi)-V_{\sigma s}\\\nonumber &&\longrightarrow
\ddot\theta+2\frac{\ddot\sigma}{\dot\sigma}\dot\theta+V_{\sigma
s}=\frac{1}{\dot\sigma^2}(\ddot\chi V_\vp-\ddot\vp
V_\chi)=3H\frac{1}{\dot\sigma^2}(\dot\vp V_\chi-\dot\chi
V_\vp)=-3H\dot\theta\hspace{4cm} QED,
\end{eqnarray}
where in above calculation we used Klein-Gordon equations
$\ddot\vp+3H\dot\vp+V_\vp=0$, $\ddot\chi+3H\dot\chi+V_\chi=0$ and
$\ddot\sigma+3H\dot\sigma+V_\sigma=0$ due to
$\dot\sigma^2=\dot\vp^2+\dot\chi^2$.
To compare the second equation
in (\ref{eq-mo-pert}) and (48) in \cite{gordon} it should be checked
that if the coefficients of $\dot{\delta\sigma}$ and $\delta\sigma$
are same or not. The former is trivial and the latter results in
checking
$2\dot\theta\frac{\ddot\sigma}{\dot\sigma}=-(\ddot\theta+3H\dot\theta+V_{\sigma
s})$. This equality has been already checked in the above. So the
result is exactly compatible with \cite{gordon}.

%\newpage

%=============================================================

\end{document}